\documentstyle[amssymb,twocolumn,aps]{revtex}

\begin{document}
\draft
\title{Vortex states in superconducting rings}
\author{B. J. Baelus, F. M. Peeters\cite{peeters} and V. A. Schweigert\cite
{schweigert}}
\address{Departement Natuurkunde, Universiteit Antwerpen (UIA), Universiteitsplein 1,%
\\
B-2610 Antwerpen, Belgium}
\date{\today}
\maketitle

\begin{abstract}
The superconducting state of a thin superconducting disk with a hole is
studied within the non-linear Ginzburg-Landau theory in which the
demagnetization effect is accurately taken into account. We find that the
flux through the hole is not quantized, the superconducting state is
stabilized with increasing size of the hole for fixed radius of the disk,
and a transition to a multi-vortex state is found if the disk is
sufficiently large. Breaking the circular summetry through a non central
location of the hole in the disk favours the multi-vortex state.
\end{abstract}


\section{Introduction}

Thanks to recent progress in microfabrication and measurement techniques, it
is possible to study the properties of superconducting samples with sizes
comparable to the penetration depth $\lambda $\ and the coherence length $%
\xi $. The properties of these mesoscopic systems are considerably
influenced by confinement effects. Therefore, the vortex state will depend
on the size and the geometry of the sample.

In the present paper, we study the properties and the vortex states of
superconducting thin disks with a hole. In the past, two limiting cases were
studied: the thin-wire loop and the disk without a hole. In 1962 Little and
Parks studied a thin-wire loop in an axial magnetic field \cite{littleparks}%
. The $T_{c}-H$ phase diagram showed a periodic component. Each time a flux
quantum $\phi _{0}=hc/2e$ penetrates the system, $T_{c}(H)$ exhibits an
oscillation. Berger and Rubinstein \cite{berger} studied mesoscopic
superconducting loops using the non-linear Ginzburg-Landau (GL) theory. They
assumed that the induced magnetic field can be neglected for samples with
sufficiently small thickness. In the limit of thin loops, the transition
between states with different angular momentum $L$ (also called vorticity)
occurs when the enclosed flux $\phi $ equals $\left( L+1/2\right) \phi _{0}$ 
\cite{thinkham}. The superconducting disk was studied by Schweigert {\it et
al} \cite{schweigPRB57,schweigPRL79,schweigPRL81,deo} (see also Ref.~\cite
{palacios}) by solving the two GL equations self-consistently. Although the
GL equations were derived to describe superconductivity near the critical
point, this theory turns out to be valid over a much broader range of
magnetic field and temperature \cite{deo,dolan}. They found that the finite
thickness of the disk influences the magnetic field profile, i.e. the
magnetic pressure, and this changes the size of the Meissner effect which is
different from the well-studied cylinder geometries \cite{moshchalkov}. The
reverse problem, i.e. the anti-dot, was studied by Bezryadin {\it et al }
\cite{bezryadin}. They obtained a phase diagram of a thin superconducting
film with a circular hole in an axial magnetic field by solving numerically
the nonlinear GL equations in the limit of a thin film. Here we generalize
the results of Ref.~\cite{schweigPRL79} to a thin circular superconducting
disk containing a circular hole.

The intermediate case of finite width loops was studied previously by
Bardeen \cite{bardeen} within the London theory. He showed that in tubes of
very small diameter and with wall thickness of the order of the penetration
depth the flux through the tube is quantized in units of $\nu \phi _{0}$
where $\nu <1$ depends on the dimensions of the system. Arutunyan and
Zharkov \cite{zharkovboek,arutunyan} found that the flux through the
effective area $\pi \left( \rho ^{\ast }\right) ^{2}$ is equal to $m\phi
_{0},$ where the effective radius $\rho ^{\ast }$ is approximately equal to
the geometric mean square of the inner radius $R_{i}$ and the outer radius $%
R_{o}$ of the cylinder; i.e. $\rho ^{\ast }=\left( R_{i}R_{o}\right) ^{1/2}$%
.\ Recently, Fomin {\it et al }\cite{fomin}{\it \ }studied square loops with
leads attached to it and found inhomogeneous Cooper pair distributions in
the loop with enhancements near the corners of the square loop. Bruyndoncx 
{\it et al }\cite{bruyndoncx} investigated infinitely thin loops of finite
width. In this case, the magnetic field induced by the supercurrents can be
neglected and the total magnetic field equals the external applied magnetic
field. Furthermore, they used the linearized GL equation which is only valid
near the superconductor/normal boundary where the density of the
superconducting condensate $\left| \psi \right| ^{2}$ is small. Only the
giant vortex state with a definite angular momentum $L$ was studied and they
concentrated on the two- (2D) to three-dimensional (3D) crossover. Berger
and Rubinstein \cite{berger2} also studied infinitely thin loops of finite
width using the non-linear GL theory and they also neglected the induced
field.

It is well known that for type-II $\left( \kappa >1/\sqrt{2}\right) $
superconductors the triangular Abrikosov vortex lattice is energetically
favored in the range $H_{c1}<H<H_{c2}$. Since the effective London
penetration depth $\Lambda =\lambda ^{2}/d$ increases considerably in thin
samples and for $d\ll \lambda $ one would expect the appearance of the
Abrikosov multi-vortex state even in thin rings made with a material with $%
\kappa <1/\sqrt{2}.$ Similar as for the case of a thin disk \cite
{schweigPRL81}, we expect that the structure of a finite number of vortices
should differ from a simple triangular arrangement in the case of a thin
ring and we expect that they will resemble the configurations found for
Coulomb clusters which are confined into a ring \cite{irina}.

In the present paper we consider circular flat disks of non-zero width with
a circular hole in it, which is not necessary in the center of the disk. The
superconducting properties are also studied deep inside the superconducting
state where: i) nonlinear effects are important, i.e. $\left| \psi \right| $
is not necessarily small, and the nonlinear GL equations have to be solved,
ii) the total magnetic field is not homogeneous, i.e. it is spatial varying
due to the Meissner effect and the flux quantization condition which may
enhance or diminish the magnetic field through the hole as compared to the
applied magnetic field, and iii) due to nonlinear effects the circular
symmetric giant vortex states are not necessarily the lowest energy states
and the magnetic field can penetrate the superconductor through single
vortices creating a multi-vortex state.

The paper is organized as follows: In Section II we present the theoretical
model. In Section III we consider a small superconducting disk with a hole
in the center. In this case we find that only the giant vortex state
appears. We study the influence of the radius of the hole on the
superconducting state. For such a small system the relation between the
local magnetic field, the current density and the Cooper pair density is
investigated and the quantization of the flux through the hole is
investigated. Next, in Section IV, we consider the case of a larger
superconducting disk with a hole in the center. For increasing magnetic
field, we find re-entrant behaviour; i.e. transition from the giant vortex
state to the multi-vortex state and back to the giant vortex state before
superconductivity is destroyed. Finally, in Section V, we investigate the
influence of the position of the hole on the vortex configuration; i.e. what
happens if we break the axial symmetry? Our results are summarized in
Section VI.

\section{Theoretical formalism}

In the present paper, we consider superconducting disks with radius $R_{o}$
and thickness $d$ with a hole inside with radius $R_{i}$ which is placed a
distance $a$ away from the center of the disk (Fig.~\ref{conf}). These
superconducting `rings' are immersed in an insulating medium with a
perpendicular uniform magnetic field $H_{0}$. To solve this problem we
follow the numerical approach of Schweigert and Peeters \cite{schweigPRB57}.
For thin disks $(d\ll \xi ,\lambda )$ they found that it is allowed to
average the GL equations over the disk thickness. Using dimensionless
variables and the London gauge $div\overrightarrow{A}=0$ for the vector
potential $\overrightarrow{A}$, we write the system of GL equations in the
following form 
\begin{equation}
\left( -i\overrightarrow{\nabla }_{2D}-\overrightarrow{A}\right) ^{2}\Psi
=\Psi \left( 1-\left| \Psi \right| ^{2}\right) ,  \label{lijn1}
\end{equation}
\begin{equation}
-\Delta _{3D}\overrightarrow{A}=\frac{d}{\kappa ^{2}}\delta \left( z\right) 
\overrightarrow{j}_{2D},  \label{lijn2}
\end{equation}
where 
\begin{equation}
\overrightarrow{j}_{2D}=\frac{1}{2i}\left( \Psi ^{\ast }\overrightarrow{%
\nabla }_{2D}\Psi -\Psi \overrightarrow{\nabla }_{2D}^{\ast }\Psi \right)
-\left| \Psi \right| ^{2}\overrightarrow{A},  \label{lijn3}
\end{equation}
is the density of superconducting current. The superconducting wavefunction
satisfies the following boundary conditions 
\begin{eqnarray}
\left. \left( -i\overrightarrow{\nabla }_{2D}-\overrightarrow{A}\right) \Psi
\right| _{\overrightarrow{r}=\overrightarrow{R_{i}}} &=&0,  \eqnum{4a}
\label{bound1} \\
\left. \left( -i\overrightarrow{\nabla }_{2D}-\overrightarrow{A}\right) \Psi
\right| _{\overrightarrow{r}=\overrightarrow{R_{o}}} &=&0,  \eqnum{4b}
\label{bound2}
\end{eqnarray}
and $\overrightarrow{A}=\frac{1}{2}H_{0}\rho \overrightarrow{e}_{\phi }$ far
away from the superconductor. Here the distance is measured in units of the
coherence length $\xi $, the vector potential in $c${\it 
h\hskip-.2em\llap{\protect\rule[1.1ex]{.325em}{.1ex}}\hskip.2em%
}$/2e\xi $, and the magnetic field in $H_{c2}=c${\it 
h\hskip-.2em\llap{\protect\rule[1.1ex]{.325em}{.1ex}}\hskip.2em%
}$/2e\xi ^{2}=\kappa \sqrt{2}H_{c}$. The ring is placed in the plane $(x,y),$
the external magnetic field is directed along the $z$-axis, and the indices $%
2D$, $3D$ refer to two- and three-dimensional operators, respectively.

The giant vortex state in a circular configuration is characterized by the
total angular momentum $L$ through $\Psi =\psi \left( \rho \right) \exp
(iL\phi ),$ where $\rho $ and $\phi $ are the cylindrical coordinates. $L$
is the winding number and gives the vorticity of the system. An arbitrary
superconducting state is generally a mixture of different angular harmonics $%
L$ due to the non-linearity of the GL equations. Nevertheless, we can
introduce an analog to the total angular momentum $L$\ which is still a good
quantum number and which is in fact nothing else then the number of vortices
in the system.

To solve the system of Eqs.(\ref{lijn1}-\ref{lijn2}), we generalized the
approach of Ref.~\cite{schweigPRB57} for disks to our fat ring
configuration. We apply a finite-difference representation of the order
parameter and the vector potential on a uniform Cartesian space grid (x,y),
with typically 128 grid points over a distance $2R_{o}$ (i.e. the diameter
of the ring), and use the link variable approach \cite{kato}, and an
iteration procedure based on the Gauss-Seidel technique to find $\Psi $. The
vector potential is obtained with the fast Fourier transform technique where
we set $\overrightarrow{A}_{\left| x\right| =R_{S},\left| y\right|
=R_{S}}=H_{0}\left( x,-y\right) /2$ at the boundary of a larger space grid $%
\left( \text{typically }R_{S}=4R_{o}\right) $.

To find the different vortex configurations, which include the metastable
states, we search for the steady-state solutions of Eqs.(\ref{lijn1}-\ref
{lijn2}) starting from different randomly generated initial conditions. Then
we increase/decrease slowly the magnetic field and recalculate each time the
exact vortex structure. We do this for each vortex configuration in a
magnetic field range where the number of vortices stays the same. By
comparing the dimensionless Gibbs free energies of the different vortex
configurations 
\begin{equation}
F=V^{-1}\int_{V}\left[ 2\left( \overrightarrow{A}-\overrightarrow{A}%
_{0}\right) \cdot \overrightarrow{j}_{2d}-\left| \Psi \right| ^{4}\right] d%
\overrightarrow{r},  \eqnum{5}
\end{equation}
where integration is performed over the sample volume $V,$ and $%
\overrightarrow{A}_{0}$ is the vector potential of the uniform magnetic
field, we find the ground state. The dimensionless magnetization, which is a
direct measure of the expelled magnetic field from the sample, is defined as 
\begin{equation}
M=\frac{\left\langle H\right\rangle -H_{0}}{4\pi },  \eqnum{6}
\label{magnform}
\end{equation}
where $H_{0}$ is the applied magnetic field. $\left\langle H\right\rangle $
is the magnetic field averaged over the sample or detector surface area $S$.

\section{Small rings: the giant vortex state}

First we discuss a small superconducting ring. Although the system is
circular symmetric in general, we are not allowed to assume that $\Psi
\left( \overrightarrow{\rho }\right) =F\left( \rho \right) e^{iL\theta }$
because of the nonlinear term in the GL-equations (\ref{lijn1}-\ref{lijn2}).
Nevertheless, we found that for sufficiently small rings the confinement
effects are dominant and this imposes a circular symmetry on the
superconducting condensate. If we can assume axial symmetry, the dimensions
of the GL equations will be reduced and thus the accuracy and the
computation time will be improved. Therefore we will use this method for
small rings. For a disk, this kind of solution with fixed $L$ has been
called ${\it 2D}${\it \ solution }by Deo {\it et al }\cite{deo} in contrast
to the ${\it 3D}${\it \ solution} of the general problem.

The ground state free energy $F$ of a superconducting disk with radius $%
R_{o}=2.0\xi $ and thickness $d=0.005\xi $ and GL parameter $\kappa =0.28$
is shown in Fig.~\ref{energy} for a hole in the center with radius $%
R_{i}/\xi =0.0$, $0.5$, $1.0$ and $1.5$, respectively. The situation with $%
R_{i}=0$ corresponds to the situation of a superconducting disk without a
hole which was already studied in Ref. \cite{schweigPRL79}. With increasing
hole radius $R_{i}$, we find that the superconducting/normal transition
(this is the magnetic field where the free energy equals zero) shifts
appreciably to higher magnetic fields and more transitions between different 
$L$-states are possible before superconductivity disappears. Furthermore,
the thin dotted curve gives the free energy for a thicker ring with
thickness $d=0.1\xi $ and $R_{i}=1.0\xi $. In comparison with the previous
results for $d=0.005\xi $, the free energy becomes more negative, but the
transitions between the different $L$-states occur almost at the same
magnetic fields. Thus increasing the thickness of the ring increases
superconductivity which is a consequence of the smaller penetration of the
magnetic field into the superconductor.

Experimentally, using magnetization measurements one can investigate the
effect of the geometry and the size of the sample on the superconducting
state. One of the main advantages of these measurements, in comparison with
resistivity measurements, is that they are non invasive and no conduction
probes are needed which may distort the superconducting state. To
investigate the magnetization of a single superconducting disk Geim {\it et
al} \cite{geim1}\ used sub-micron Hall probes. The Hall cross acts like a
magnetometer, where in the ballistic regime\ the Hall voltage is determined
by the average magnetic field through the Hall cross region \cite{li}.
Hence, by measuring the Hall resistance, one obtains the average magnetic
field, and consequently, the magnetic field expelled from the Hall cross
(Eq.~(\ref{magnform})),\ which is a measure for the\ magnetization of the
superconductor \cite{deo}. As in the case of a superconducting disk \cite
{deo}, the field distribution in the case of thin superconducting rings is
extremely non-uniform inside as well as outside the sample and therefore the
detector size will have an effect on the measured magnetization. To
understand this effect of the detector, we calculate the magnetization for a
superconducting ring with outer radius $R_{o}=2.0\xi ,$ thickness $%
d=0.005\xi $ and two values of the inner radius $R_{i}=0.5\xi $ and $%
R_{i}=1.5\xi $ by averaging the magnetic field (see Eq.~(\ref{magnform}))
over several detector sizes $S$. The results are shown in Fig.~\ref
{magnetization}(a,b) as a function of the applied magnetic field for $%
R_{i}=0.5\xi $ and $R_{i}=1.5\xi $, respectively. The solid curve (curve
(i)) shows the calculated magnetization if we average the magnetic field
over the superconducting volume, and curve (ii) is the magnetization after
averaging the field over a circular area with radius $R_{o}$, i.e.
superconductor and hole. Notice that in the Meissner regime, i.e. $%
H_{0}/H_{c2}<0.7$ where $L=0,$ the magnetization of superconductor + hole is
larger than the one of the superconductor alone which is due to flux
expulsion from the hole. For $L\geq 1$ the reverse is true because now flux
is trapped in the hole. Experimentally, one usually averages the magnetic
field over a square Hall cross region. Therefore, we calculated the
magnetization by averaging the magnetic field over a square region
(curve(iii))\ with width equal to $2R_{o},$ i.e. the diameter of the ring.
Curve (iv) shows the magnetization if the sides of the square detector are
equal to $(2+1/2)R_{o}$. Increasing the size of the detector, decreases the
observed magnetization because the magnetic field is averaged over a larger
region which brings $\left\langle H\right\rangle $ closer to the applied
field $H_{0}$

Having the free energies of the different giant vortex configurations for
several values of the hole radius varying from $R_{i}=0.0\xi $ to $%
R_{i}=1.8\xi $, we construct an equilibrium vortex phase diagram. Fig.~\ref
{phaseri} shows this phase diagram for a superconducting disk with radius $%
R_{o}=2.0\xi $, thickness $d=0.005\xi $ and for $\kappa =0.28$. The solid
curves indicate where the ground state of the free energy changes from one $L
$-state to another and the thick solid curve gives the
superconducting/normal transition. The latter exhibits a small oscillatory
behavior which is a consequence of the Little-Parks effect.\ Notice that the
superconducting/normal transition is moving to larger fields with increasing
hole radius $R_{i}$ and more and more flux can be trapped. In the limit $%
R_{i}\rightarrow R_{o}$, the critical magnetic field is infinite and there
are an infinite number of $L$-states possible which is a consequence of the
enhancement of surface conductivity for very small samples \cite
{schweigPRB57,surface}. Because of the finite grid, we were not able to
obtain accurate results for $R_{i}\approx R_{o}$. The dashed lines connect
our results for hole radius $R_{i}=1.8\xi $ with the results for $%
R_{i}\rightarrow R_{o}$ \cite{thinkham}, where the transitions between the
different $L$-states occur when the enclosed flux is $\phi =(L+1/2)\phi _{0}$%
, where $\phi _{0}=ch/2e$ is the elementary flux quantum. Notice that for
rings of non zero width, i.e. $R_{i}\neq R_{o},$ the $L\rightarrow L+1$
transition occurs at higher magnetic field than predicted from the condition 
$\phi =(L+1/2)\phi _{0}$. The discrepancy increases with increasing width of
the ring and with increasing $L$. Starting from $R_{i}=0$ we find that with
increasing $R_{i}$ the Meissner state disappears at smaller $H_{0}.$ The
hole in the center of the disk allows for a larger penetration of the
magnetic field which favors the $L=1$ state. This is the reason why the $%
L=0\rightarrow L=1$ transition moves to a lower external field while the $%
L=1\rightarrow L=2$ transition initially occurs for larger $H_{0}$ with
increasing $R_{i}$. When the hole size becomes of the order of the width of
one vortex the $L=1\rightarrow L=2$ transition starts to move to lower
fields and the $L=2$ state becomes more favorable.

The effect of the thickness of the ring on the phase diagram is investigated
in Fig.~\ref{phased}. The thin solid curves indicate where the ground state
of the free energy changes from one $L$-state to another one, while the
thick curve gives the superconducting/normal transition. Notice that the
transition from the $L=0$ to the $L=1$ state and the superconducting/normal
transition depends weakly on the thickness of the ring. For increasing
thickness $d,$ the $L=2$ state becomes less favorable and disappears for $%
d\gtrsim 0.7\xi .$ In this case, there is a transition from the $L=1$ state
directly to the normal state. Increasing $d$ stabilizes the different $L$%
-states up to larger magnetic fields. This is due to the increased expulsion
of the applied field from the superconducting ring \cite{schweigPRB57}.

In the next step, we investigate three very important and mutually dependent
quantities: the local magnetic field $H$, the Cooper pair density $\left|
\Psi \right| ^{2}$ and the current density $j$. We will discuss these
quantities as a function of the radial position $\rho .$ For this study we
distinguish two situations, i.e. $R_{i}\ll R_{o}$ and $R_{i}\lesssim R_{o}.$
In the first case the sample behaves more like a superconducting disk, and
in the second case like a superconducting loop.

First, we consider a superconducting disk with radius $R_{o}=2.0\xi $ and
thickness $d=0.1\xi $ with a hole with radius $R_{i}=0.5\xi $ in the center.
The free energy and the magnetization (after averaging over the
superconductor+hole) for such a ring are shown in Fig.~\ref{enR05}. The
dashed curves give the free energy and the magnetization for the different $%
L $-states, and the solid curve is the result\ for the ground state. Fig.~%
\ref{magnR05}(a,b) shows the local magnetic field $H$, Fig.~\ref{magnR05}%
(c,d) the Cooper pair density $\left| \Psi \right| ^{2}$, and Fig.~\ref
{magnR05}(e,f) the current density $j$ as function of the radial position $%
\rho $ for such a ring at the $L$-states and magnetic fields as indicated by
the open circles in Fig.~\ref{enR05}(a,b).

For low magnetic field, the system is in the $L=0$ state, i.e. the Meissner
state, and the flux trapped in the hole is considerably suppressed. Hence,
the local magnetic field inside the hole is lower than the external applied
magnetic field as is shown in Fig.~\ref{magnR05}(a) by curves 1 and 2.
Please\ notice that the plotted magnetic field is scaled by the applied
field $H_{0}$.\ In the $L=0$ state the superconductor expels the magnetic
field by inducing a supercurrent which tries to compensate the applied
magnetic field in the superconductor and inside the hole. This is called the
diamagnetic Meissner effect. As long as $L$ equals zero, the induced current
has only to compensate the magnetic field at the outside of the ring and,
therefore, the current flows in the whole superconducting material in the
same direction and the size increases with increasing field. This is clearly
shown in Fig.~\ref{magnR05}(e) by curves 1 and 2 where the current density $%
j $ becomes more negative for increasing $H_{0}/H_{c2}$. Notice also that
the current density is more negative at the outside than at the inside of
the superconducting ring which leads to a stronger depression of the Cooper
pair density at the outer edge as compared to the inner edge of the ring
(see curves 1 and 2 in Fig.~\ref{magnR05}(c)).

At $H_{0}/H_{c2}=0.745$, the ground state changes from the $L=0$ to the $L=1$
state (see Fig.~\ref{enR05}(a)). Suddenly more flux becomes trapped in the
hole (compare curve 1 with curve 3 in Fig.~\ref{magnR05}(a)), the local
magnetic field inside the hole increases and becomes larger than the
external magnetic field $H_{0}$. In the $L=1$ state, there is a sharp peak
in the magnetic field\ at the inner boundary because of demagnetization
effects. Consequently, more current is needed to compensate the magnetic
field near the inner boundary than near the outer boundary (see curve 3 in
Fig.~\ref{magnR05}(e)). The sign of the current near the inner boundary
becomes positive (the current direction reverses), but the sign of the
current near the outer boundary does not change.\ This can be explained as
follows: Near the inner boundary $\left( \rho \gtrsim 1.0\xi \right) $ the
magnetic field is compressed into the hole (paramagnetic effect), while near
the outer boundary $\left( \rho \lesssim 2.0\xi \right) $ the magnetic field
is expelled to the insulating environment (diamagnetic effect). The sign
reversal of $j$ occurs at $\rho =\rho ^{\ast }$ and later we will show that
the flux through the circular area with radius $\rho ^{\ast }$ is exactly
quantized. At the $L=0$ to the $L=1$ transition the maximum in the Cooper
pair density (compare curves 1 and 3 in Fig.~\ref{magnR05}(c)) shifts from $%
\rho =R_{i}$ to $\rho =R_{o}$. Further increasing the external field
increases the Cooper pair density near the inner boundary initially (compare
curve 3 and 4 in Fig.~\ref{magnR05}(c)), because the flux in the hole has to
be compressed less. The point $\rho ^{\ast }$, where $j=0$, shifts towards
the inner boundary of the ring. Further increasing the external magnetic
field, the Cooper pair density starts to decrease (see curves 5 and 6 in
Fig.~\ref{magnR05}(d)) and attains its maximum near the inner boundary. The
current near the inner boundary becomes less positive (see curves 5 and 6 in
Fig.~\ref{magnR05}(f)), i.e. less shielding of the external magnetic field
inside the hole (see curves 5 and 6 in Fig.~\ref{magnR05}(b)), and near the
outer boundary $j$ becomes less negative which shields the magnetic field
from the superconductor+hole. Thus at the outer edge the local magnetic
field has a local maximum which decreases with applied magnetic field $H_{0}$%
.

At $H_{0}/H_{c2}\approx 2.0325$, the ground state changes from the $L=1$
state to the $L=2$ state and extra flux is trapped in the hole. The changes
in the magnetic field distribution, the Cooper pair density and the current
density are analogous to the changes at the first transition. For example,
the magnetic field inside the hole increases compared to the external
magnetic field (curve 7 in Fig.~\ref{magnR05}(b)), the radius $\rho ^{\ast }$
increases substantially (curve 7 in Fig.~\ref{magnR05}(f)) and the maximum
in the Cooper pair density shifts to the outer boundary (curve 7 in Fig.~\ref
{magnR05}(d)).

For $R_{i}\ll R_{o}$ and $L>0$, the superconducting state consists of a
combination of the paramagnetic and the diamagnetic Meissner state, like for
a disk. For $R_{i}\lesssim R_{o}$ we expect that the sample behaves like a
loop and, hence, the superconducting state is a pure paramagnetic Meissner
state or a pure diamagnetic Meissner state.

We consider a superconducting disk with radius $R_{o}=2.0\xi $ and thickness 
$d=0.1\xi $ with a hole with radius $R_{i}=1.8\xi $ in the center. The free
energy and the magnetization (after averaging over the superconductor+hole)\
for such a ring are shown in Fig.~\ref{enR18}. The dashed curves give the
free energy and the magnetization for the different $L$-states, and the
solid curve is the result\ for the ground state. Fig.~\ref{magnR18}(a,b)
shows the local magnetic field $H$, and Fig.~\ref{magnR18}(c,d) the current
density $j$ as function of the radial position $\rho $ for such a ring at
the $L$-states and magnetic fields as indicated by the open circles in Fig.~%
\ref{enR18}(a,b). The Cooper pair density has almost no structure and is
practically constant over the ring and will, therefore, not be shown.

For $L=0$, the situation is the same as for $R_{i}\ll R_{o}$. The magnetic
field is expelled from the superconductor and the hole to the outside of the
system, i.e. diamagnetic Meissner effect. The current flows in the whole
superconducting material in the same direction (curves 1 and 2 in Fig.~\ref
{magnR18}(a)) and the size increases with increasing external field $H_{0}$
(curves 1 and 2 in Fig.~\ref{magnR18}(c)). At $H_{0}/H_{c2}=0.27$, the
ground state changes from the $L=0$ state to the $L=1$ state and suddenly
more flux is trapped in the hole. The local magnetic field inside the hole
becomes larger than the external field $H_{0}$ and there is a sharp peak
near the inner boundary (curves 3 and 4 in Fig.~\ref{magnR18}(a)). In
contrast to the situation for $R_{i}\ll R_{o},$ there is no peak near the
outer boundary, which means that the magnetic field is only expelled to the
hole, i.e. paramagnetic Meissner effect. The induced current flows in the
reverse direction in the whole superconductor (curves 3 and 4 in Fig.~\ref
{magnR18}(c)). For increasing external magnetic field, the magnetic field
inside the hole, the height of the demagnetization peak and hence the size
of the current decrease (see curves 3 and 4 in Fig.~\ref{magnR18}(a,c)).
Further increasing the field, the superconducting state transforms into a
diamagnetic Meissner state. The magnetic field is now expelled to the
outside of the sample (curves 5 and 6 in Fig.~\ref{magnR18}(b)) and the
direction of the current is the same everywhere in the ring (curves 5 and 6
in Fig.~\ref{magnR18}(d)). At $H_{0}/H_{c2}=0.82,$ the ground state changes
from the $L=1$ state to the $L=2$ state. The changes in the magnetic field
distribution (see curve 7 in Fig.~\ref{magnR18}(b)) and the current density
(see curve 7 in Fig.~\ref{magnR18}(c)) are analogous to the changes at the
first transition. The diamagnetic state transforms into a paramagnetic state.

For a narrow ring with finite width, the superconductor is in the
paramagnetic or the diamagnetic Meissner state, like for a superconducting
loop. Contrary to this infinitely narrow ring case, for narrow finite width
rings the superconducting state can also consist of a combination of these
two states, i.e. the direction of the supercurrent in the inner part of the
ring is opposite to the outer part.

Now we will investigate the flux quantization in the fat ring. Fig.~\ref
{fluxhole} shows the flux through the hole (a) and through the
superconductor+hole (b) as function of the applied magnetic field $H_{0}$
for a superconducting disk with radius $R_{o}=2.0\xi $ and thickness $%
d=0.1\xi $ with a hole in the center with radius $R_{i}=1.0\xi $. The dashed
curves show the flux for the different $L$-states, the solid curve for the
ground state and the thin solid line is the flux through the hole if the
sample is in the normal state. It is apparent that the flux through the hole
(or through the superconducting ring + hole) is {\it not quantized }(see
Fig.~\ref{fluxhole}(a)). At $H_{0}/H_{c2}=0.4575$ suddenly more flux enters
the hole and the ground state\ changes from the $L=0$ state to the $L=1$
state. At this transition also the flux increase through the hole is not
equal to one flux quantum $\phi _{0}$. It is generally believed that the
flux through a superconducting ring is quantized. But as was shown in Ref.~ 
\cite{groff} this is even no longer true for hollow cylinders when the
penetration length is larger than the thickness of the cylinder wall. The
present result is a generalization of this observation to mesoscopic ring
structures. Note that for the case of Fig.~\ref{fluxhole} the penetration
length is $\lambda /\xi =0.28$ and the effective penetration length $\Lambda
/\xi =0.78$ is comparable to the width of the ring $\left(
R_{o}-R_{i}\right) /\xi =1.0$.

For $L>0,$ the superconducting current equals zero at a certain `effective'
radial position $\rho =\rho ^{\ast }$. It is the flux through the circular
area with radius $\rho ^{\ast }$ which is quantized and not necessarily the
flux through the hole of our disk. Following Ref. \cite{zharkovboek} we
integrate the supercurrent $j$\ over a closed contour $C$ lying entirely
inside the superconducting material which embraces the opening, one finds 
\begin{equation}
\oint_{C}\left( \frac{4\pi }{c}\kappa ^{2}\xi ^{2}\overrightarrow{j}+%
\overrightarrow{A}\right) \cdot d\overrightarrow{l}=L\phi _{0}.
\label{flux1}
\end{equation}
Starting from the definition of flux and using Stokes' theorem, the flux can
be written as 
\begin{equation}
\phi =\int \overrightarrow{H}\cdot d\overrightarrow{S}=\int rot%
\overrightarrow{A}\cdot d\overrightarrow{S}=\oint_{C}\overrightarrow{A}\cdot
d\overrightarrow{l}\text{.}  \label{flux2}
\end{equation}
If the contour $C$ is chosen in such a way that the current $j=$ $j_{S}=0$
on this contour, then the flux through the surface area bounded by this
contour is quantized 
\begin{equation}
\phi =\oint_{C}\overrightarrow{A}\cdot d\overrightarrow{l}=L\phi _{0}.
\label{flux3}
\end{equation}

To demonstrate that this is indeed true we show in Fig.~\ref{firho}(a) the
current density as a function of the radial position $\rho $ and in Fig.~\ref
{firho}(b) the flux through a circular area with radius $\rho $ for a
superconducting disk with radius $R_{o}=2.0\xi $ and thickness $d=0.1\xi $
with a hole in the center with radius $R_{i}=1.0\xi $ in the presence of an
external magnetic field $H_{0}/H_{c2}=1.6075$ for the case of three
different giant vortex states; i.e.$\ L=1,2,3$. In the $L=1$ state, the
current density equals zero at a distance $\rho ^{\ast }/\xi \approx 1.16$
from the center and the flux through an area with this radius is exactly
equal to one flux quantum $\phi _{0}$. For $L=2$ and $L=3$, the current
density $j$\ equals zero at $\rho ^{\ast }/\xi \approx 1.56$ and $1.91$,
respectively, and the flux through the area with this radius $\rho ^{\ast }$
is exactly equal to $2\phi _{0}$ and $3\phi _{0}$, respectively. We find
that $\rho ^{\ast }$ depends on the external applied magnetic field and on
the value of $L$, contrary to the results of Arutunyan and Zharkov \cite
{zharkovboek,arutunyan} who found that the effective radius $\rho ^{\ast }$
is approximately equal to the geometric mean square of the inner radius $%
R_{i}$ and the outer radius $R_{o}$ of the cylinder; i.e. $\rho ^{\ast
}=\left( R_{i}R_{o}\right) ^{1/2},$ which for the case of Fig.~\ref{rhoster}%
(a) would give $\rho ^{\ast }=1.41\xi $. The results of Refs.~\cite
{zharkovboek,arutunyan} were obtained within the London limit.\ The
dependence of $\rho ^{\ast }$ as function of the applied magnetic field is
shown in Fig.~\ref{rhoster}(a). The dashed curves give the $\rho ^{\ast }$
of the different $L$-states. For increasing magnetic field and fixed $L$,
the value of $\rho ^{\ast }$ decreases, i.e. the ''critical'' radius moves
towards the inner boundary. The solid curve gives $\rho ^{\ast }$ for the
ground state. At the $L\rightarrow L+1$ transitions, $\rho ^{\ast }$ jumps
over a considerable distance towards the outside of the superconducting
ring, the size of the jumps decreases with increasing $L$. The dotted lines
in the figure give the two boundaries of the superconducting ring: the outer
boundary $R_{o}=2.0\xi $ and the inner boundary $R_{i}=1.0\xi $. Remark that
in Fig.~\ref{rhoster} there is no $\rho ^{\ast }$ given for the $L=0$ state,
because only the external magnetic field has to be compensated so that the
current has the same sign everywhere inside the ring and there exists no $%
\rho ^{\ast }$. In Fig.~\ref{rhoster}(b) we repeated this calculation for a
hole with radius $R_{i}=1.5\xi $ where superconductivity remains to higher
magnetic fields and many more $L\rightarrow L+1$ transitions are possible.
The result of Refs.~\cite{zharkovboek,arutunyan} gives in this case $\rho
^{\ast }=1.73\xi $. Remark that the results for the $L=1$ and the $L=2$
states are not connected. The reason is that just before the $L=1\rightarrow
L=2$ transition the critical current for the $L=1$ state is strictly
positive in the whole ring and hence $\rho ^{\ast }$ is not defined. Notice
that the results in Fig.~\ref{rhoster}(a,b) oscillate around the average
value $\rho ^{\ast }=\sqrt{R_{i}R_{o}}$ as given by Refs.~\cite
{zharkovboek,arutunyan}.

For even narrower rings the current in the ring is mostly diamagnetic or
paramagnetic and the region in magnetic field over which both directions of
current occur in the ring becomes very narrow. This is illustrated in Fig.~%
\ref{flux18} for a ring with outer radius $R_{o}=2.0\xi $ and inner radius $%
R_{i}=1.8\xi $. We plot in Fig.~\ref{flux18}(a) the flux through the hole
which becomes very close to the external flux, i.e. the flux without any
superconductor. In Fig.~\ref{flux18}(b) the value of the current at the
inner and the outer side of the ring is shown which illustrates nicely that
over large ranges of the magnetic field the current in the ring flows in one
direction. The free energy becomes minimum (see Fig.~\ref{flux18}(c)) at a
magnetic field where the current in the rings flows in both directions.

\section{Large rings: the Multi-vortex state}

Until now, we considered only small rings. In such rings, the confinement
effect dominates and we found that only the giant vortex states are stable
and possible multi-vortex states have always larger energies if they exist.
Now we will consider larger superconducting disks in which multi-vortex
states can nucleate for certain magnetic fields. As an example we take a fat
ring with outer radius $R_{o}=4.0\xi $, thickness $d=0.005\xi ,$ $\kappa
=0.28$ and for different values of the inner radius. Fig.~\ref{energymulti}%
(a) shows the free energy for such a ring with inner radius $R_{i}=0.4\xi $
as a function of the applied magnetic field $H_{0}$. The different giant
vortex states are given by the thin solid curves and the multi-vortex state
by the thick solid curves. The open circles indicate the transitions from
the multi-vortex state to the giant vortex state. In this ring, multi-vortex
states exist with winding number $L=3$ up to $7$. For $L=3,4,5$
multi-vortices occur both as metastable states as well as in the ground
state, while for $L=6,7$ they are only found in the metastable state. Notice
there is no discontinuity in the free energy at the transitions from the
multi-vortex state to the giant vortex state for fixed winding number $L$.
Fig.~\ref{energymulti}(b) shows the magnetization $M$ for this ring as a
function of $H_{0}$ after averaging the field $H$ over the superconducting
ring (without the hole). The dashed curves give the results for the
different $L$-states, the thin solid curve for the ground state, the thick
solid curves for the multi-vortex states and the open circles indicate the
transition from the multi-vortex state to the giant vortex states. Notice
that the latter transitions are smooth, there are no discontinuities in the
magnetization.

Now, we investigate the flux $\phi $ through the hole for the $L=4$
multi-vortex and giant vortex state for the case of the above ring. Fig.~\ref
{fluxmulti} shows the flux $\phi $ through a circular area of radius $\rho $
for different values of the applied magnetic field $H_{0}$. Curves 1,2,3 and
4 are the results for $H_{0}/H_{c2}=0.72$, $0.795$, $0.87$ (i.e.
multi-vortex states) and $0.945$ (i.e. giant vortex state), respectively.
There is no qualitative difference between the 4 curves, i.e. no qualitative
difference between the multi-vortex states and the giant vortex state. In
the inset, we show the flux through the hole with radius $R_{i}=0.4\xi $ and
through the superconductor+hole as a function of the applied magnetic field
for a fixed value of the winding number, i.e. $L=4$. The solid circles
indicate the magnetic fields considered in the main figure and the open
circle indicates the position of the transition from multi-vortex state to
giant vortex state. The slope of the curves increases slightly at the
applied magnetic field, where there is a transition form the multi-vortex
state to the giant vortex state. This agrees with the result for a disk \cite
{schweigPRL81} that the giant$\leftrightarrow $multi-vortex transition is a
second-order phase transition.

The Cooper pair density $\left| \psi \right| ^{2}$\ for the previous four
configurations is shown in Fig.~\ref{densityR4L4}. The darker the region,
the larger the density and thus vortices are given by white regions. At the
magnetic field $H_{0}/H_{c2}=0.72$ (Fig.~\ref{densityR4L4}(a)) we see
clearly three multi-vortices. With increasing magnetic field, these
multi-vortices start to overlap and move to the center (Fig.~\ref
{densityR4L4}(b,c)) and finally they combine to one giant vortex in the
center (Fig.~\ref{densityR4L4}(d)) Please notice that a theory based on the
London limit will not be able to give such a complicated behaviour, because
in such a theory vortices are rather point-like objects.

The free energies of the different vortex configurations were calculated for
different values of the hole radius which we varied from $R_{i}=0$ to $%
R_{i}=3.6\xi $. From these results we constructed an equilibrium vortex
phase diagram. First, we assumed axial symmetry, where only giant vortex
states occur and the order parameter is given by $\Psi \left( 
\overrightarrow{\rho }\right) =F\left( \rho \right) e^{iL\theta }$. In the
phase diagram (Fig.~\ref{fasemulti}) the solid curves separate the regions
with different number of vortices (different $L$-states). In the limit $%
R_{i}\rightarrow 0$ we find the previous results of Ref.~\cite{schweigPRL81}
for a superconducting disk. The radius of the giant vortex $R_{L}$ in the
center of the disk increases with increasing $L$, because it has to
accommodate more flux, i.e. $R_{L}/\xi \sim \sqrt{L/(H_{0}/H_{c2})}$. Hence,
if we make a little hole in the center of the disk, this will not influence
the $L\rightarrow L+1$ transitions as long as $R_{i}\ll R_{L}$ as is
apparent from Fig.~\ref{fasemulti}. For sufficient large hole radius $R_{i}$%
, the hole starts to influence the giant vortex configuration and the
magnetic field needed to induce the $L\rightarrow L+1$ transition increases.
For example, the transition field from the $L=7$ state to the $L=8$ state
reaches its maximum for a hole radius $R_{i}\approx 2.0\xi $ which occurs
for $H_{0}/H_{c2}\approx 1.5$. The above rough estimate gives $R_{L}/\xi
\sim 2.16$ which is very close to $R_{i}/\xi \sim 2.0.$ Further increasing $%
R_{i}$, the hole becomes so large that more and more flux is trapped inside
the hole, and consequently a smaller field is needed to induce the $%
L\rightarrow L+1$ transition. Because of finite grid size, we were limited
to $R_{i}\leq 3.6\xi $. The results we find for $R_{i}=3.6\xi $ are
extrapolated to $\phi =(L+1/2)\phi _{0}$ for $R_{i}=R_{o}$. The thick curve
in Fig.~\ref{fasemulti} gives the superconducting/normal transition. For low
values of $R_{i}$, this critical magnetic field is independent of $R_{i}$,
because the hole is smaller than the giant vortex state in the center and
hence the hole does not influence the superconducting properties near the
superconducting/normal transition. For $R_{i}\gtrsim 2.0\xi $, this field
starts to increase drastically. Therefore, more and more $L$-states appear.
In the limit $R_{i}\rightarrow R_{o}$, the critical magnetic field is
infinite and there are an infinite number of $L$-states possible which is a
consequence of the enhancement of surface conductivity for very small
superconducting samples \cite{surface}.

Next, we consider the general situation where the order parameter is allowed
to be a mixture of different giant vortex states and thus we no longer
assume axial symmetry of the superconducting wavefunction. We found that the
transitions between the different $L$-states are not influenced by this
generalization, but that for certain magnetic fields the ground state is
given by the multi-vortex state instead of the giant vortex state. In Fig.~%
\ref{fasemulti} the shaded regions correspond to the multi-vortex states and
the dashed curves are the boundaries between the multi-vortex and the giant
vortex states. For $L=1$, the single vortex state and the giant vortex state
are identical. In the limit $R_{i}\rightarrow 0$, the previous results of
Ref.~\cite{schweigPRL81} for a superconducting disk are recovered. For
increasing hole radius $R_{i}$, the $L=2$ multi-vortex state disappears as a
ground state for $R_{i}>0.15\xi $. If $R_{i}$ is further increased, the
ground state for $L=5$ up to $L=9$ changes from giant vortex state to
multi-vortex state and again to giant vortex state. For example, for $%
R_{i}=2.0\xi $ the multi-vortex state exists only in the $L=9$ state. Notice
that for small $R_{i}$ the region of multi-vortex states increases and
consequently the hole in the center of the disk stabilizes the multi-vortex
states, at least for $L>2$. For large $R_{i},$ i.e. narrow rings, the giant
vortex state is the energetic favorable one because confinement effects
start to dominate which impose the circular symmetry on $\psi .$ For fixed
hole radius $R_{i}\leq 2.0\xi $ and increasing magnetic field we find always
at least one transition from giant vortex state to multi-vortex state and
back to giant vortex state (re-entrant behaviour).\ Remark that the ground
state for $L\geq 10$ is in the giant vortex state irrespective of the value
of the magnetic field. Near the superconducting/normal transition the
superconducting ring is in the giant vortex state because now
superconductivity exists only near the edge of the sample and consequently
the superconducting state will have the same symmetry as the outer edge of
the ring and thus it will be circular symmetric.

Finally, for $L\geq 3$ the multi-vortex state not necessarily consists of $L$
vortices in the superconducting material. Often they consist of a
combination of a big vortex trapped in the hole and some multi-vortices in
the superconducting material. This is clearly shown in Fig.~\ref
{contourmulti} where a contour plot of the local magnetic field is given for
a superconducting disk with radius $R_{o}=4.0\xi $ and thickness $d=0.005\xi 
$ with a hole in the center with radius $R_{i}=0.6\xi $ (Fig.~\ref
{contourmulti}(a)) and $R_{i}=1.0\xi $ (Fig.~\ref{contourmulti}(b)). As
usual we took $\kappa =0.28$. The dashed thick circle is the outer radius,
while the small solid black circle is the inner radius of the ring. Low
magnetic fields are given by light regions and darker regions indicate
higher magnetic fields. In this way, multi-vortices in the superconducting
area are dark spots. In Fig.~\ref{contourmulti}(a) the local magnetic field
is shown for an applied magnetic field $H_{0}/H_{c2}=0.895$. Although the
winding number is $L=4$, there are only 3 vortices in the superconducting
material and one vortex appears in the hole in the center of the disk. This
is clearly shown in Fig.~\ref{faseR4}(a) where the phase $\varphi $ of the
order parameter is shown along different circular loops $\overrightarrow{r}%
\rightarrow Ce^{i\chi }$ inside the superconductor. The solid curve gives
the phase near the outer edge of the ring $(C=3.95\xi )$ and the dashed
curve near the inner edge of the ring $(C=0.7\xi )$. When encircling the
ring, the phase difference $\Delta \varphi $ in the first case is $4$ times $%
2\pi ,$ while in the second case it is $\Delta \varphi =1\times 2\pi $. The
phase difference is always given by $\Delta \varphi =L\times 2\pi $, with $L$
the winding number. In Fig.~\ref{contourmulti}(b) a contour plot of the
local magnetic field is shown for a ring with $R_{i}=1.0\xi $ which leads to 
$L=6$ at $H_{0}/H_{c2}=1.145$. Only 4 vortices are in the superconducting
material and one giant vortex in the center (partially in the hole) with $%
L=2 $. Fig.~\ref{faseR4}(b) shows the phase $\varphi $ of the order
parameter for $C=3.95\xi $ (solid curve) and $1.2\xi $ (dashed curve), where
the phase differences are $\Delta \varphi =6\times 2\pi $ and $2\times 2\pi
, $ respectively. Notice that the flux $\phi $ through the hole equals $\phi
\approx 0.19\phi _{0}$ for the case of Fig.~\ref{contourmulti}(a) and $\phi
\approx 0.36\phi _{0}$ for the case of Fig.~\ref{contourmulti}(b) and is
thus not equal to a multiple of the flux quantum $\phi _{0}.$

\section{Non-symmetric geometry}

So far, we investigated the influence of the size of the hole on the vortex
configuration for superconducting rings of different sizes. We found that
for small rings, only the axially symmetrical situation occurs, i.e. the
giant vortex states. For large rings the multi-vortex state can be
stabilized for certain values of the magnetic field. In the next step, we
purposely break the axial symmetry by moving the hole away from the center
of the superconducting disk over a distance $a$.

As an example, we consider a superconducting disk with radius $R_{o}=2.0\xi $
and thickness $d=0.005\xi $ with a hole with radius $R_{i}=0.5\xi $ moved
over a distance $a=0.6\xi $ in de $x$-direction. Fig.~\ref{energieverpl}
shows the free energy and magnetization (defined through the field expelled
from the superconducting ring without the hole) as function of the magnetic
field. The solid curve indicates the ground state, while the dashed curves
indicate the metastable states for increasing and decreasing field. The
vertically dotted lines separate the regions with different winding number$\
L$. Notice that hysteresis is only\ found for the first transition from the $%
L=0$ to the $L=1$ state and not for the higher transitions which are
continuous. At the transition from the $L=1$ state to the $L=2$ state and
further to the $L=3$ state, the free energy and the magnetization vary
smoothly (see Fig.~\ref{energieverpl}(a,b)). In the insets of Fig.~\ref
{energieverpl}(a), we show the Cooper pair density for such a sample at
magnetic field $H_{0}/H_{c2}=0.145$, $1.02,$ $2.145$ and $2.52$,
respectively, where the ground state is given by a state with $L=0,1,2,3$
respectively. High Cooper pair density is given by dark regions, while light
regions indicate low Cooper pair density. For $H_{0}/H_{c2}=0.145$, we find
a high Cooper pair density in the entire superconducting ring. There is
almost no flux trapped in the circular area with radius smaller than $R_{o}$%
. After the first transition at $H_{0}/H_{c2}\approx 0.75$, suddenly more
flux is trapped in the hole which substantially lowers the Cooper pair
density in the superconductor. Notice that the trapped flux tries to restore
the circular symmetry in the Cooper pair density and that the density of the
superconducting condensate is largest in the narrowest region of the
superconductor. The next inset shows the Cooper pair density of the $L=2$
state where an additional vortex appears. Some flux is passing through the
hole (i.e. winding number is one around the hole), while one flux line is
passing through the superconducting ring and a local vortex (the normal
region with zero Cooper pair density) is created. In the $L=3$ state
superconductivity is destroyed in part of the sample which contains flux
with winding number $L=2$ and the rest of the flux passes through the hole.
Hence, by breaking the circular symmetry of the system, multi-vortex states
are stabilized. Remember that for the corresponding symmetric system, i.e. $%
a=0$, only giant vortex states were found.

Having the magnetic fields for the different $L\rightarrow L+1$ transitions
for superconducting rings with different positions of the hole, i.e.
different values of $a,$ we constructed the phase diagram shown in Fig.~\ref
{phasediagrama}. The thin curves (solid curves when the magnetization is
discontinuous and dashed curves when the magnetization is continuous)
indicate the magnetic field at which the transition from the $L$-state to
the $L+1$-state occurs, while the thick solid curve gives the
superconducting/normal transition.

In order to show that the stabilization of the multi-vortex state due to an
off-center hole is not peculiar for $R_{o}=2.0\xi $, we repeated the
previous calculation for a larger superconducting disk with radius $%
R_{o}=5.0\xi $ and thickness $d=0.005\xi $ containing a hole with radius $%
R_{i}=2.0\xi $. In Fig.~\ref{contourverpl}(a) the Cooper pair density is
shown for such a system with the hole in the center, while in Fig.~\ref
{contourverpl}(b) the hole is moved away form the center over a distance $%
a=1.0\xi $ in the negative $y$-direction. The externally applied magnetic
field is the same in both cases, $H_{0}/H_{c2}=0.77$, which gives a winding
number $L=4$ and $L=5$ for the ground state of the symmetric and the
non-symmetric geometry, respectively. The assignment of the winding number
can be easily checked from Figs.~\ref{contourverpl}(c,d) which show contour
plots of the corresponding phase of the superconducting wavefunction. If the
hole is at the center of the disk, the ground state is a giant vortex state.
Moving the center of the hole to the position $\left( x/\xi ,y/\xi \right)
=\left( 0,-1\right) $, two vortices appear in the superconducting material
while the hole contains three vortices. Notice that in this case, although
the magnetic field is kept the same and the amount of superconducting
material is not altered, changing the symmetry of the system alters the
winding number.

\section{Conclusion}

In conclusion, we studied the superconducting state of thin superconducting
disks with a hole. The effect of the size and the position of the hole on
the vortex configuration was investigated. For small superconducting disks
with a hole in the center, only giant vortex states exist and for increasing
hole radius $R_{i}$ more and more $L$-states occur before the superconductor
becomes normal. For larger superconducting disks with a hole in the center,
we found multi-vortex states in a certain magnetic field range. For certain
fixed hole radius, and for increasing magnetic field, the giant vortex state
changes into a multi-vortex state and back into the giant-vortex state
(re-entrant behaviour) before superconductivity is destroyed. Near the
superconducting/normal transition and for a narrow superconducting ring
(i.e. $R_{i}\approx R_{o}$) we always found the giant vortex state as the
ground state irrespective of the size, thickness and width of the ring. The
effect of the position of the hole, i.e. decreasing the symmetry of the
system, was also investigated. Moving the hole off-center: 1) can transform
the $L\rightarrow L+1$ transition into a continuous one, 2) the stability of
metastable states is strongly reduced, 3) it favours the multi-vortex state
even for small disks, and 4) the winding number $L$ can increase even at a
fixed magnetic field.

The flux through the hole is {\it not} quantized. We were able to define an
effective ring size $\rho ^{\ast }$ such that inside this ring the flux is
exactly quantized. The value of $R_{i}\leq \rho ^{\ast }\leq R_{o}$ depends
on $L$ and is an oscillating function of the magnetic field. For narrow
rings it is only possible to define such a $\rho ^{\ast }$ in narrow ranges
of the magnetic field and the flux through the hole is very close to the
applied flux. The magnetic fields from the screening currents are too small
to substantially modify the flux inside the ring. On the other hand, the
magnetic field increment $\Delta H$ or the flux increase $\Delta \phi =\pi
R_{o}^{2}\Delta H$ to induce the $L\rightarrow L+1$ transition is only
quantized for narrow rings. This is illustrated in Fig.~\ref{deltafi} in
case of $R_{o}=4.0\xi $ for different values of the inner radius. Notice
that for $R_{i}\ll R_{o}$ we find that $\Delta \phi $ is an oscillating
function of $L$. It is substantially larger than $\phi _{0}$ for small $L$,
it is smaller than $\phi _{0}$ for intermediate $L$ and it approaches $\phi
_{0}$ from above for large $L.$ We found earlier that for $R_{i}\approx
R_{o} $ the flux through $\rho ^{\ast }=\sqrt{R_{i}R_{o}}$ is quantized in $%
\phi _{0}$ and therefore we expect $\Delta \phi ^{\ast }=\pi \left( \rho
^{\ast }\right) ^{2}\Delta H=\phi _{0}$. Our definition of $\Delta \phi $
considers the flux through the superconducting ring + hole which for the
condition $\Delta \phi ^{\ast }=\phi _{0}$ gives $\Delta \phi /\phi
_{0}\approx 6.67,2.5,1.54$ and $1.11$ for $R_{i}/\xi =0.6,1.6,2.6$ and $3.6$%
, respectively. These results for $R_{i}/\xi =2.6$ and $3.6$ agree rather
well with our numerical results presented in Fig.~\ref{deltafi}; i.e. $%
\Delta \phi /\phi _{0}\approx 1.5$ and $1.1$, respectively. The results
presented in Fig.~\ref{deltafi} agree qualitatively with the recent
theoretical results of Bruyndoncx {\it et al} \cite{bruyndoncx} who studied
rings in a homogeneous magnetic field.

\section{Acknowledgments}

This work was supported by the Flemish Science Foundation (FWO-Vl), IUAP-VI
and ESF-Vortex Matter. One of us (BJB) is supported by BOF (SFO) (Antwerp)
and FMP is a research director with the FWO-Vl. One of us (FMP) acknowledges
discussions with J. P. Lindelof.

\begin{figure}[tbp]
\caption{The configuration: a superconducting disk with radius $R_{o}$ and
thickness $d$ with a hole inside with radius $R_{i}$ which is placed a
distance $a$ away from the center.}
\label{conf}
\end{figure}

\begin{figure}[tbp]
\caption{The ground state free energy as a function of the applied magnetic
field $H_{0}$ of a superconducting disk with radius $R_{o}=2.0\protect\xi $,
thickness $d=0.005\protect\xi $ and $\protect\kappa =0.28$ for a hole in the
center with radius $R_{i}/\protect\xi =0.0$, $0.5$, $1.0$ and $1.5$
respectively. The thin dotted curve gives the free energy of a thicker ring
with thickness $d=0.1\protect\xi $ and $R_{i}=1.0\protect\xi $. The free
energy is in units of $F_{0}=H_{c}^{2}V/8\protect\pi $.}
\label{energy}
\end{figure}

\begin{figure}[tbp]
\caption{The magnetization as a function of the applied magnetic field $%
H_{0} $\ for a superconducting disk with radius $R_{o}=2.0\protect\xi $ and
thickness $d=0.005\protect\xi $ and $\protect\kappa =0.28$ for a hole in the
center with radius $R_{i}/\protect\xi =0.5$ (a) and $R_{i}/\protect\xi =1.5$
(b). Curve (i) is the calculated magnetization if we average the magnetic
field over the superconducting volume; curve (ii) after averaging the field
over the area with radius $R_{o}$, i.e. superconductor and hole; curves
(iii) and (iv) after averaging the magnetic field over a square region with
widths equal to $2R_{o}$ and $(2+1/2)R_{o}$, respectively.}
\label{magnetization}
\end{figure}

\begin{figure}[tbp]
\caption{Phase diagram: the relation between the hole radius $R_{i}$ and the
magnetic fields $H_{0}$ at which giant vortex transitions $L\rightarrow L+1$
takes place for a superconducting disk with radius $R_{o}=2.0\protect\xi $
and thickness $d=0.005\protect\xi $ and for $\protect\kappa =0.28$. The
solid curves indicate where the ground state of the free energy changes from
one $L$-state to another one and the thick solid curve gives the
superconducting/normal transition.}
\label{phaseri}
\end{figure}

\begin{figure}[tbp]
\caption{The same as Fig.$~$\ref{phaseri}$\ $but now for varying thickness $%
d $ of the ring for fixed$\ R_{o}=2.0\protect\xi ,$ $R_{i}=0.5\protect\xi $
and $\protect\kappa =0.28$.}
\label{phased}
\end{figure}

\begin{figure}[tbp]
\caption{The free energy (a)\ and the magnetization after averaging over the
superconductor+hole (b) as a function of the applied magnetic field\ for a
superconducting disk with outer radius $R_{o}=2.0\protect\xi $ and thickness 
$d=0.1\protect\xi $ with a hole with radius $R_{i}=0.5\protect\xi $ in the
center $\left( \protect\kappa =0.28\right) $ for the different $L$-states
(dashed curves) and for the ground state (solid curves). The open circles
are at: (1) $L=0$, $H_{0}/H_{c2}=0.745$; (2) $L=0$, $H_{0}/H_{c2}=0.995$;
(3) $L=1$, $H_{0}/H_{c2}=0.745$; (4) $L=1$, $H_{0}/H_{c2}=0.995$; (5) $L=1$, 
$H_{0}/H_{c2}=1.7825$; (6) $L=1$, $H_{0}/H_{c2}=2.0325$; and (7) $L=2$, $%
H_{0}/H_{c2}=2.0325$.}
\label{enR05}
\end{figure}

\begin{figure}[tbp]
\caption{The local magnetic field $H$ (a,b), the Cooper pair density $\left|
\Psi \right| ^{2}$ (c,d) and the current density $j$ (e,f) for the
situations indicated by the open circles in Fig.~\ref{enR05}\ as a function
of the radial position $\protect\rho $ for a superconducting disk with
radius $R_{o}=2.0\protect\xi $ and thickness $d=0.1\protect\xi $ with a hole
with radius $R_{i}=0.5\protect\xi $ in the center $\left( \protect\kappa
=0.28\right) $.}
\label{magnR05}
\end{figure}

\begin{figure}[tbp]
\caption{The free energy (a)\ and the magnetization after averaging over the
superconductor+hole (b) as a function of the applied magnetic field\ for a
superconducting disk with outer radius $R_{o}=2.0\protect\xi $ and thickness 
$d=0.1\protect\xi $ with a hole with radius $R_{i}=1.8\protect\xi $ in the
center $\left( \protect\kappa =0.28\right) $ for the different $L$-states
(dashed curves) and for the ground state (solid curves). The open circles
are at: (1) $L=0$, $H_{0}/H_{c2}=0.27$; (2) $L=0$, $H_{0}/H_{c2}=0.395$; (3) 
$L=1$, $H_{0}/H_{c2}=0.27$; (4) $L=1$, $H_{0}/H_{c2}=0.395$; (5) $L=1$, $%
H_{0}/H_{c2}=0.695$; (6) $L=1$, $H_{0}/H_{c2}=0.82$; and (7) $L=2$, $%
H_{0}/H_{c2}=0.82$.}
\label{enR18}
\end{figure}

\begin{figure}[tbp]
\caption{The local magnetic field $H$ (a,b) and the current density $j$
(c,d) for the situations indicated by the open circles in Fig.~\ref{enR18}\
as a function of the radial position $\protect\rho $ for a superconducting
disk with radius $R_{o}=2.0\protect\xi $ and thickness $d=0.1\protect\xi $
with a hole with radius $R_{i}=1.8\protect\xi $ in the center $\left( 
\protect\kappa =0.28\right) $.}
\label{magnR18}
\end{figure}

\begin{figure}[tbp]
\caption{The flux through (a) the hole and (b) the superconductor + the hole
as a function of the applied magnetic field $H_{0}$ for a superconducting
disk with radius $R_{o}=2.0\protect\xi $ and thickness $d=0.1\protect\xi $
with a hole in the center with radius $R_{i}=1.0\protect\xi $. The dashed
curves show the flux for the different $L$-states, the solid curve for the
ground state. The thin solid line is the flux in the absence of
superconductivity.}
\label{fluxhole}
\end{figure}

\begin{figure}[tbp]
\caption{The current density (a) and the flux (b) as a function of the
radial position $\protect\rho $ for a superconducting ring with $R_{o}=2.0%
\protect\xi ,$ $R_{i}=1.0\protect\xi $, $d=0.1\protect\xi $ and $\protect%
\kappa =0.28$ for $L=1$ (solid curves), $L=2$ (dashed curves) and $L=3$
(dotted curves) at an applied magnetic field $H_{0}/H_{c2}=1.6075$.}
\label{firho}
\end{figure}

\begin{figure}[tbp]
\caption{The dependence of the effective radius $\protect\rho ^{\ast }$ as
function of the applied magnetic field for a superconducting disk with
radius $R_{o}=2.0\protect\xi $ and thickness $d=0.1\protect\xi $ with a hole
in the center with radius $R_{i}=1.0\protect\xi $ (a), and $R_{i}=1.5\protect%
\xi $ (b). The dashed curves show $\protect\rho ^{\ast }$ for the different $%
L$-states and the solid curve is for the ground state.}
\label{rhoster}
\end{figure}

\begin{figure}[tbp]
\caption{(a) The flux through the hole of the ring as function of the
applied magnetic field. The thin dashed line is the applied flux. (b) The
current at the inner side (solid curve) and at the outer side (dashed curve)
of the ring, and (c) the free energy of the ground state as function of the
applied magnetic field.}
\label{flux18}
\end{figure}

\begin{figure}[tbp]
\caption{(a) The free energy as function of the applied magnetic field for a
superconducting ring with $R_{o}=2.0\protect\xi $, $R_{i}=0.4\protect\xi $, $%
d=0.005\protect\xi $, and $\protect\kappa =0.28$. The different giant vortex
states are shown by the thin solid curves and the multi-vortex states by the
thick solid curves. (b) The magnetization for the same sample after
averaging over the superconducting ring only. The different $L$-states are
given by the dashed curves, the ground state by the thin solid curves and
the multi-vortex states by the thick solid curves. The transitions from
multi-vortex state to giant vortex state are indicated by the open circles.}
\label{energymulti}
\end{figure}

\begin{figure}[tbp]
\caption{The flux $\protect\phi $ through a circular area of radius $\protect%
\rho $ for different values of the applied magnetic field; $H_{0}$. Curves
1,2,3 and 4 give the results for $H_{0}/H_{c2}=0.72$ (1), $0.795$ (2), $0.87$
(3) (i.e. multi-vortex states) and $0.945$ (4) (i.e. giant vortex state),
respectively. The inset shows the flux through the hole with radius $%
R_{i}=0.4\protect\xi $ and through the superconductor + hole as a function
of the applied magnetic field for a fixed value of the winding number, i.e. $%
L=4$. The solid circles indicate the magnetic fields considered in the main
figure and the open circles indicate the transition from multi-vortex to
giant vortex state. }
\label{fluxmulti}
\end{figure}

\begin{figure}[tbp]
\caption{The Cooper pair density corresponding to the four situations of
Fig.~\ref{fluxmulti}: $H_{0}/H_{c2}=0.72$ (a); $H_{0}/H_{c2}=0.795$ (b); $%
H_{0}/H_{c2}=0.87$ (c); and $H_{0}/H_{c2}=0.945$ (d). Dark regions indicate
high density, light regions low density. The thick circles indicate the
inner and the outer edge of the ring.}
\label{densityR4L4}
\end{figure}

\begin{figure}[tbp]
\caption{Equilibrium vortex phase diagram for a superconducting disk with
radius $R_{o}=4.0\protect\xi $, thickness $d=0.005\protect\xi $ with a hole
with radius $R_{i}$ in the center. The solid curves show the transitions
between the different $L$-states, the thick solid curve shows the
superconducting/normal transition and the dotted lines connect the results
for $R_{i}=3.6\protect\xi $ with the results in the limit $R_{i}\rightarrow
R_{o}$. The shaded regions indicate the multi-vortex states and the dashed
curves separate the multi-vortex states from the giant vortex states.}
\label{fasemulti}
\end{figure}

\begin{figure}[tbp]
\caption{Contour plot of the local magnetic field for a superconducting disk
with radius $R_{o}=4.0\protect\xi $ and thickness $d=0.005\protect\xi $ $%
\left( \protect\kappa =0.28\right) $ with a hole in the center with (a)
radius $R_{i}=0.6\protect\xi $ at $H_{0}/H_{c2}=0.895$ for $L=4$; and (b)
radius $R_{i}=1.0\protect\xi $ at $H_{0}/H_{c2}=1.145$ for $L=6$. The dashed
thick circle is the outer radius, the small solid thick circle the inner
radius. Low magnetic fields are given by light regions and dark regions
indicate higher magnetic fields.}
\label{contourmulti}
\end{figure}

\begin{figure}[tbp]
\caption{The phase $\protect\varphi $ of the order parameter calculated on a
circle ${\bf r}\rightarrow Ce^{i\protect\chi }$ as a function of the angle $%
\protect\chi $ for a superconducting ring with $R_{o}=4.0\protect\xi $, $%
d=0.005\protect\xi $ and $\protect\kappa =0.28$; (a) $R_{i}=0.6\protect\xi $%
, $H_{0}/H_{c2}=0.895$, $L=4$ and $C=3.95\protect\xi $ (solid curve) and $%
C=0.7\protect\xi $ (dashed curve); (b) $R_{i}=1.0\protect\xi $, $%
H_{0}/H_{c2}=1.145$, $L=6$ and $C=3.95\protect\xi $ (solid curve) and $C=1.2%
\protect\xi $ (dashed curve).}
\label{faseR4}
\end{figure}

\begin{figure}[tbp]
\caption{The free energy (a)\ and magnetization after averaging over the
superconducting ring only (b) as function of the applied magnetic field for
a superconducting disk with radius $R_{o}=2.0\protect\xi $ and thickness $%
d=0.005\protect\xi $ with a hole with radius $R_{i}=0.5\protect\xi $ moved
over a distance $a=0.6\protect\xi $ in the $x$-direction. The solid curve
indicates the ground state, the dashed curve the results for increasing and
decreasing field. The vertically dotted lines separate the regions with
different vorticity $L$. The insets show the Cooper pair density at magnetic
field $H_{0}/H_{c2}=0.145$, $1.02$, $2.145$ and $2.52$, where the ground
state is given by a state with $L=0,1,2,3$ respectively. High Cooper pair
density is given by dark regions, while light regions indicate low Cooper
pair density. }
\label{energieverpl}
\end{figure}

\begin{figure}[tbp]
\caption{The relation between the displacement $a$ and the transition
magnetic fields for a superconducting ring with $R_{0}/\protect\xi =2.0,$ $d/%
\protect\xi =0.005,$ $R_{i}/\protect\xi =0.5$ and $\protect\kappa =0.28$.
The thin solid curves give the $L\rightarrow L+1$ transitions when it
corresponds to a first-order phase transition while a dashed curve is used
when it is continuous. The thick solid curve is the superconducting/normal
transition.}
\label{phasediagrama}
\end{figure}

\begin{figure}[tbp]
\caption{The Cooper pair density for a superconducting disk with radius $%
R_{o}=5.0\protect\xi $ and thickness $d=0.005\protect\xi $ with a hole with
radius $R_{i}=2.0\protect\xi $ (a) in the center, and (b) moved away form
the center over a distance $a=1.0\protect\xi $ in the negative $y$%
-direction. The applied magnetic field is the same in both cases: $%
H_{0}/H_{c2}=0.77.$ High Cooper pair density is given by dark regions, low
Cooper pair density by light regions. The corresponding contour plot of the
phase of the superconducting wavefunction is given in (c) and (d).}
\label{contourverpl}
\end{figure}

\begin{figure}[tbp]
\caption{The flux increase $\Delta \protect\phi =\protect\pi R_{o}^{2}\Delta
H$ needed to induce the $L\rightarrow L+1$ transition for a superconducting
ring with $R_{o}=4.0\protect\xi $ for different values of the inner radius $%
R_{i}$. The interconnecting lines are a guide to the eye.}
\label{deltafi}
\end{figure}

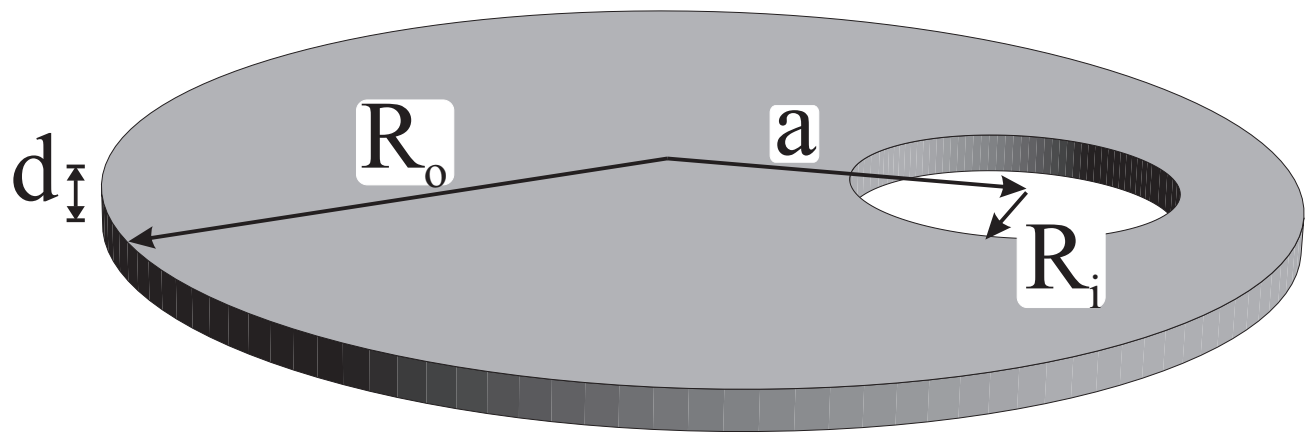
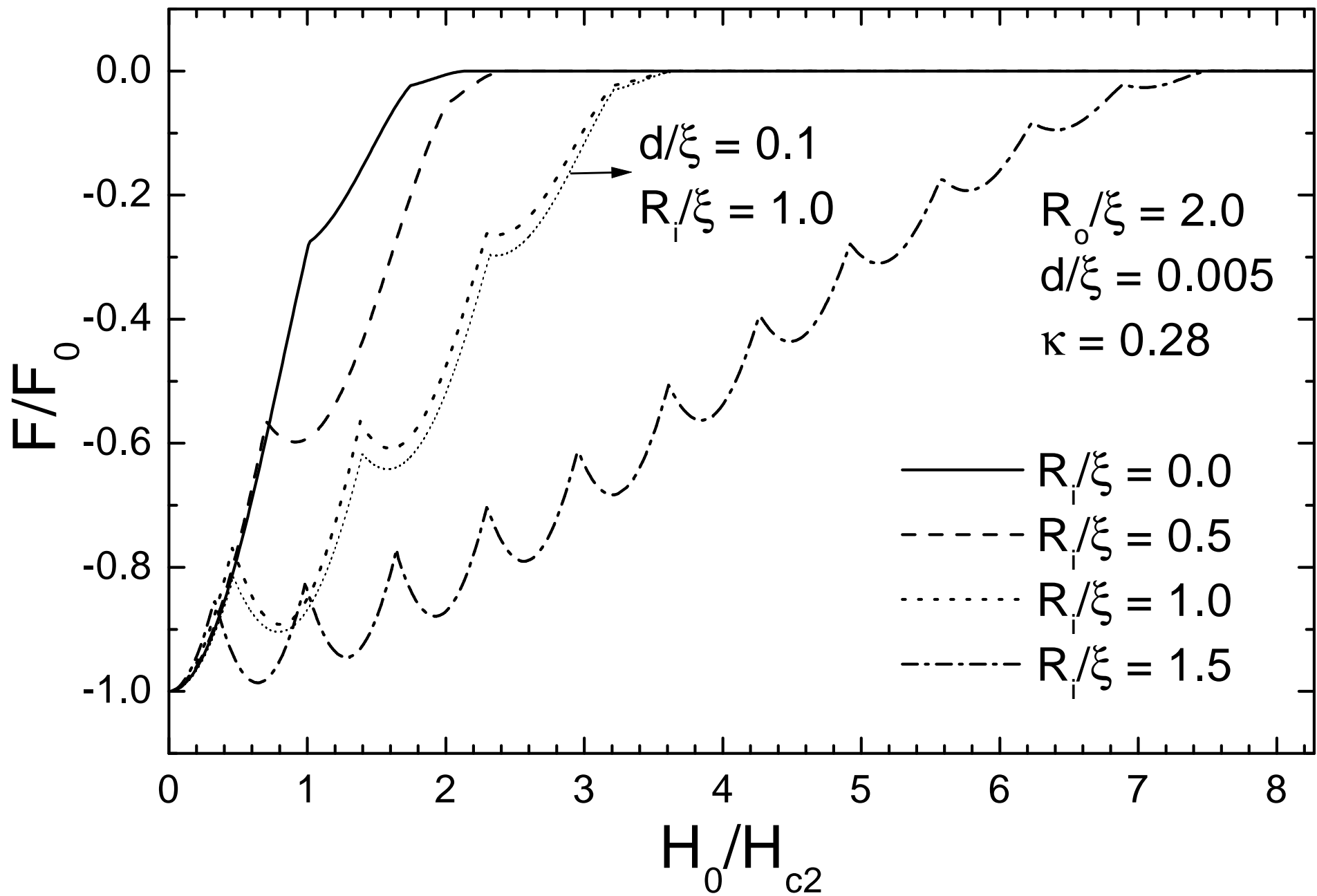
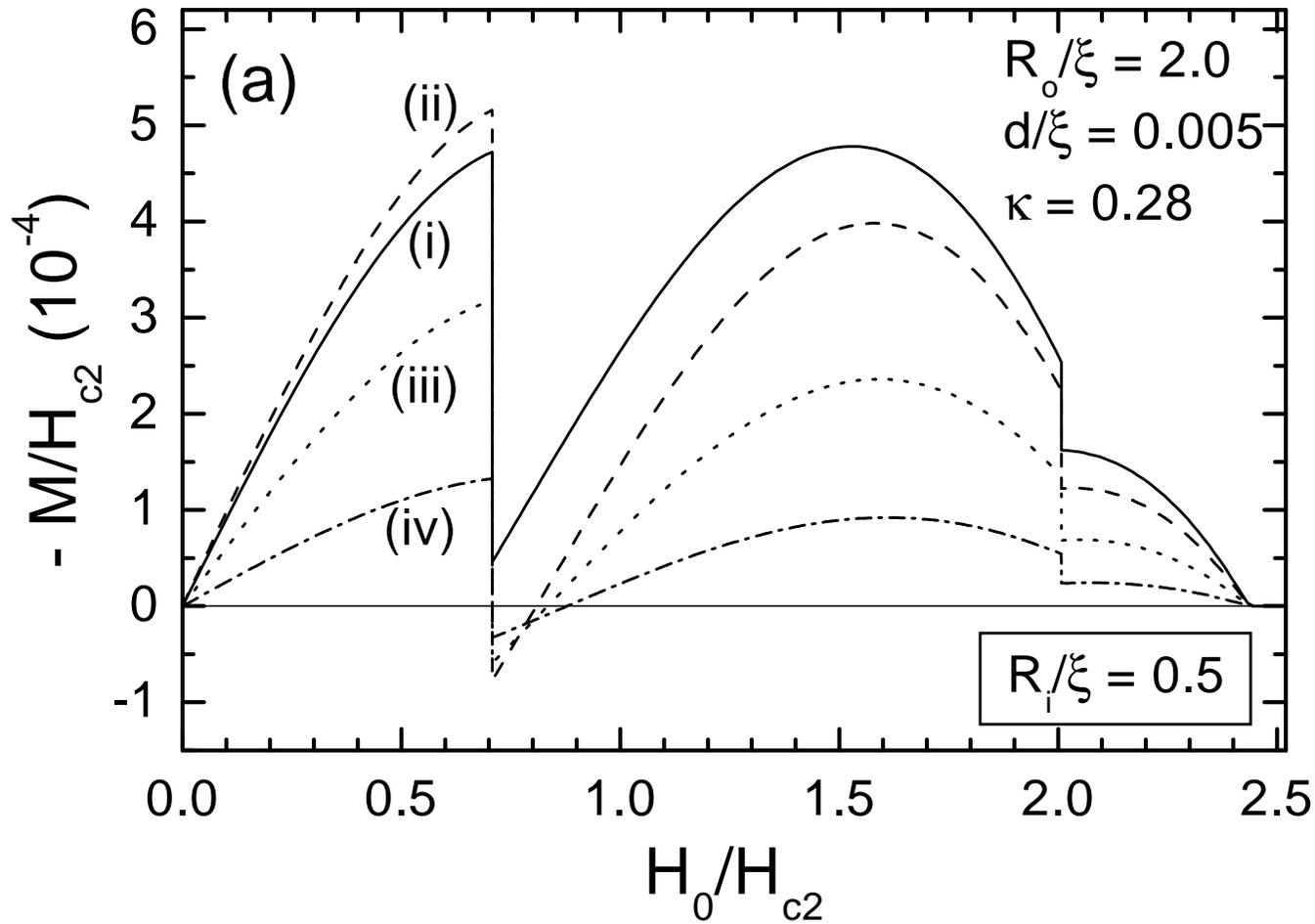
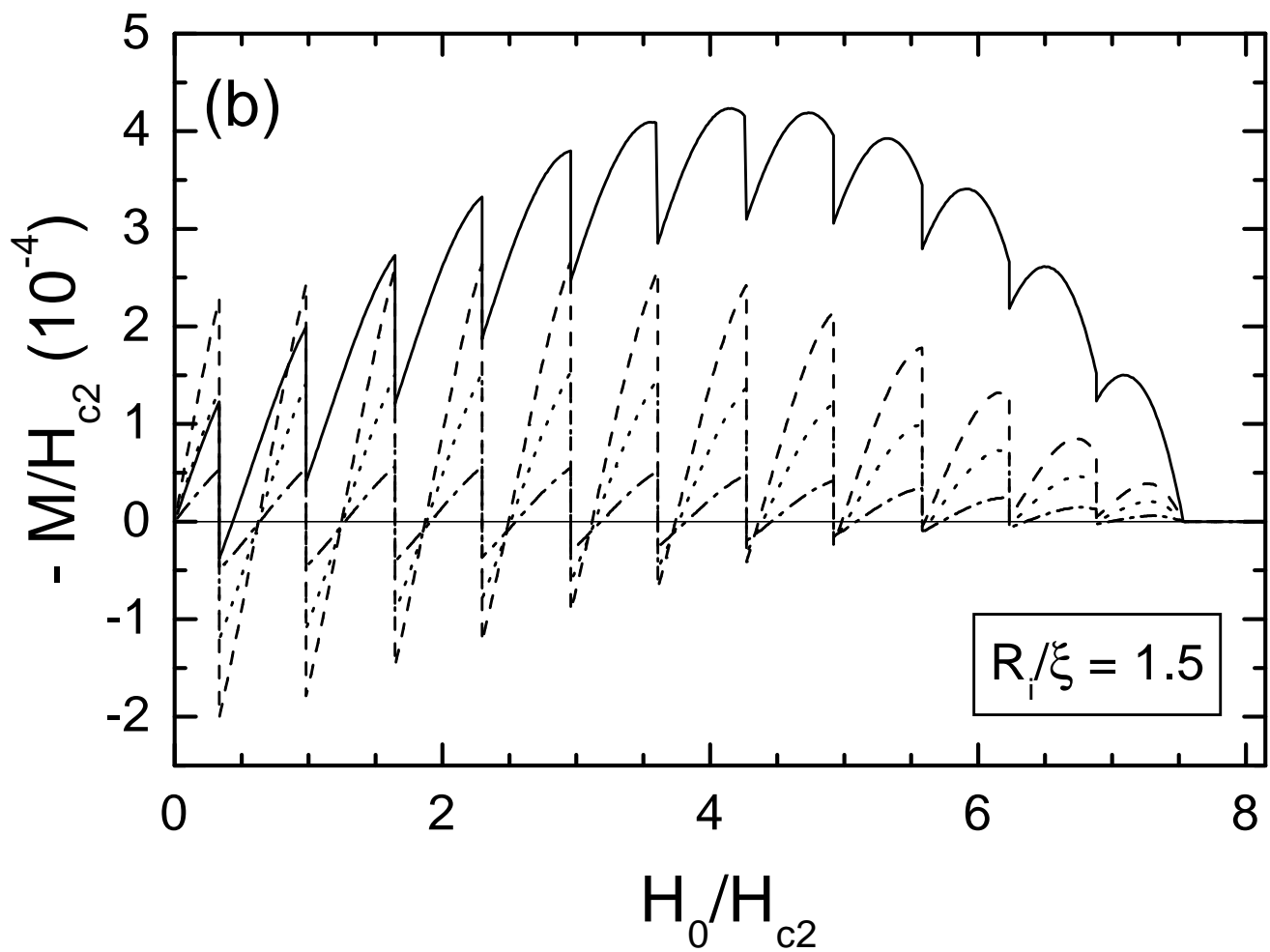
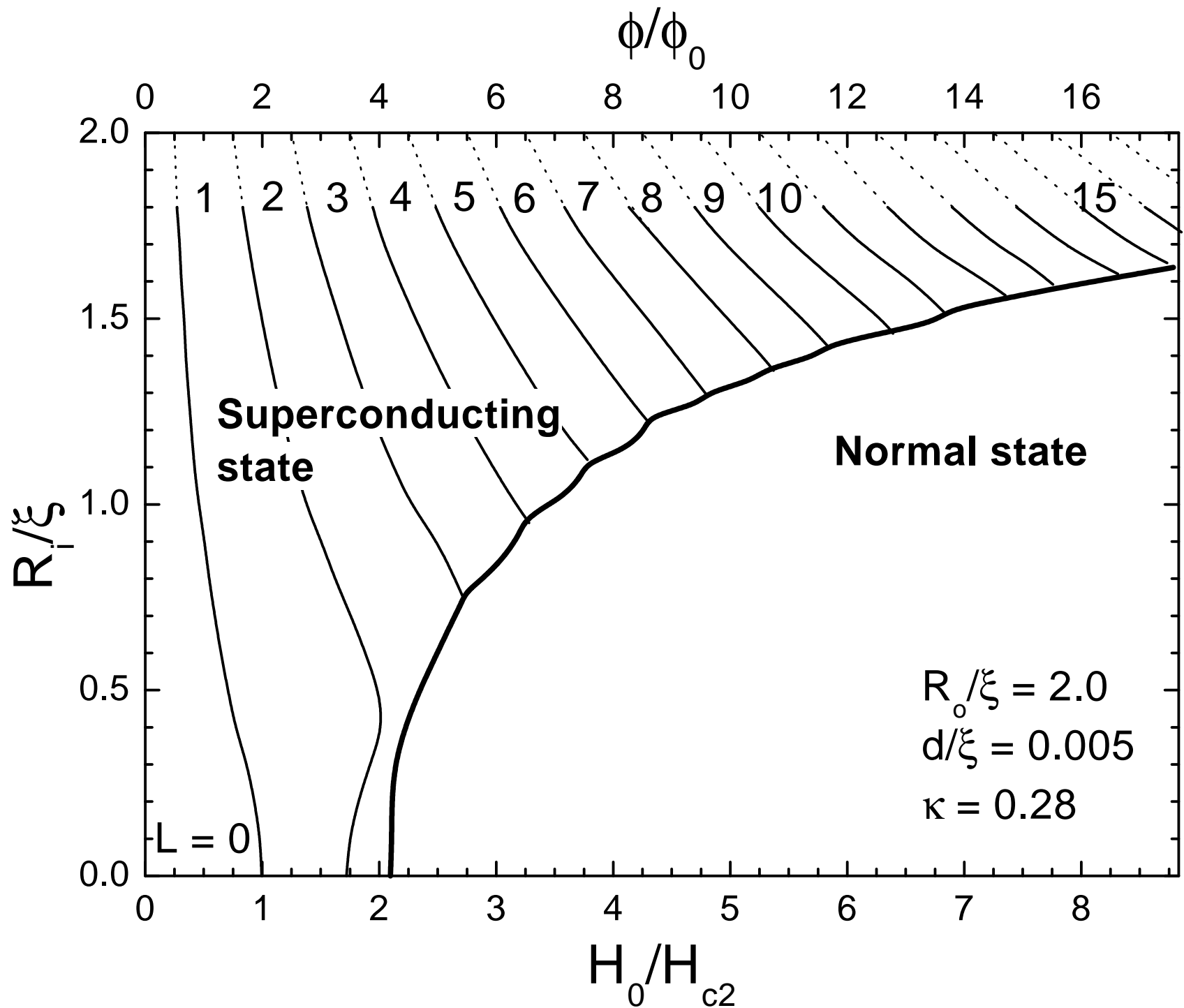
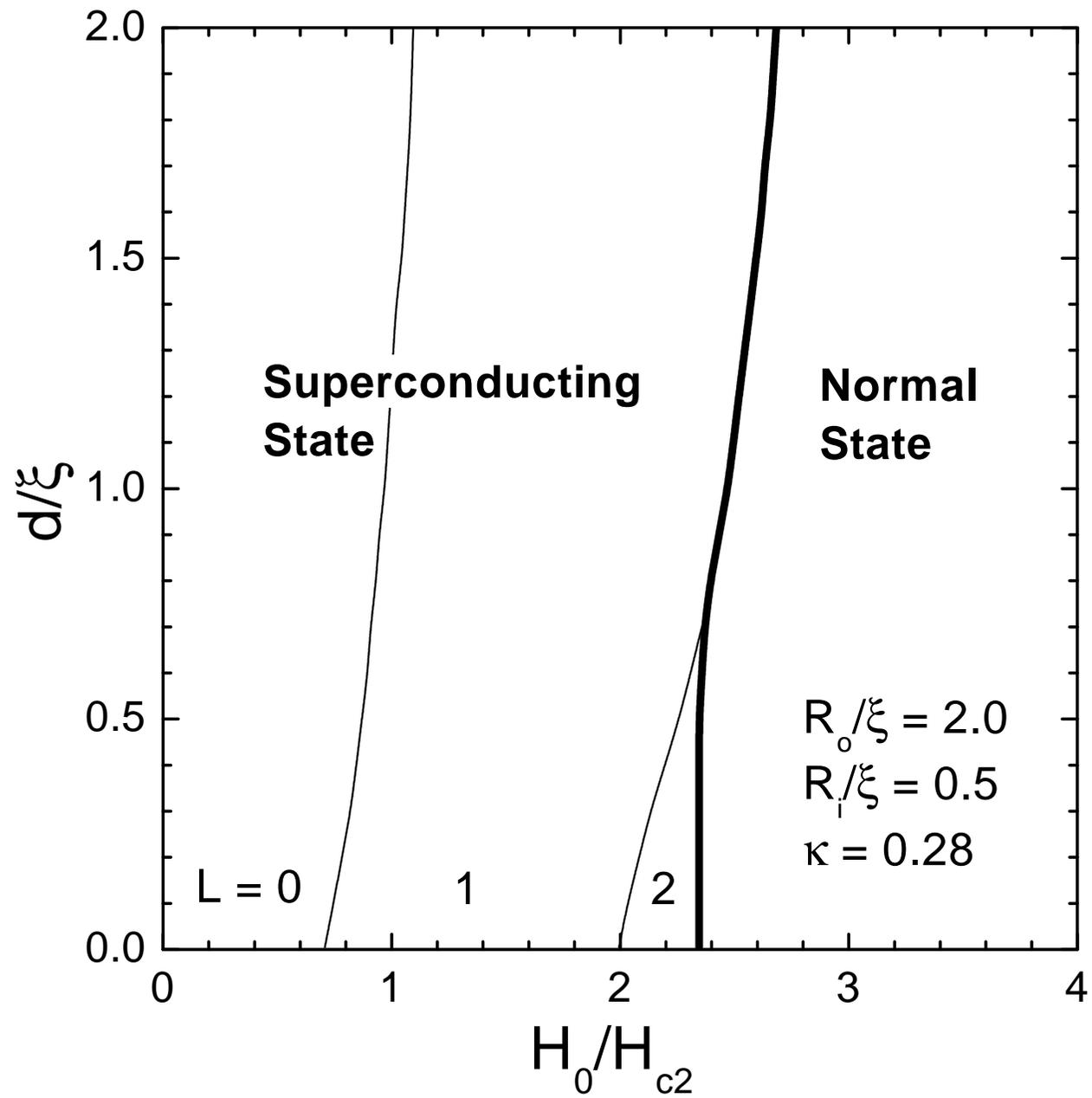
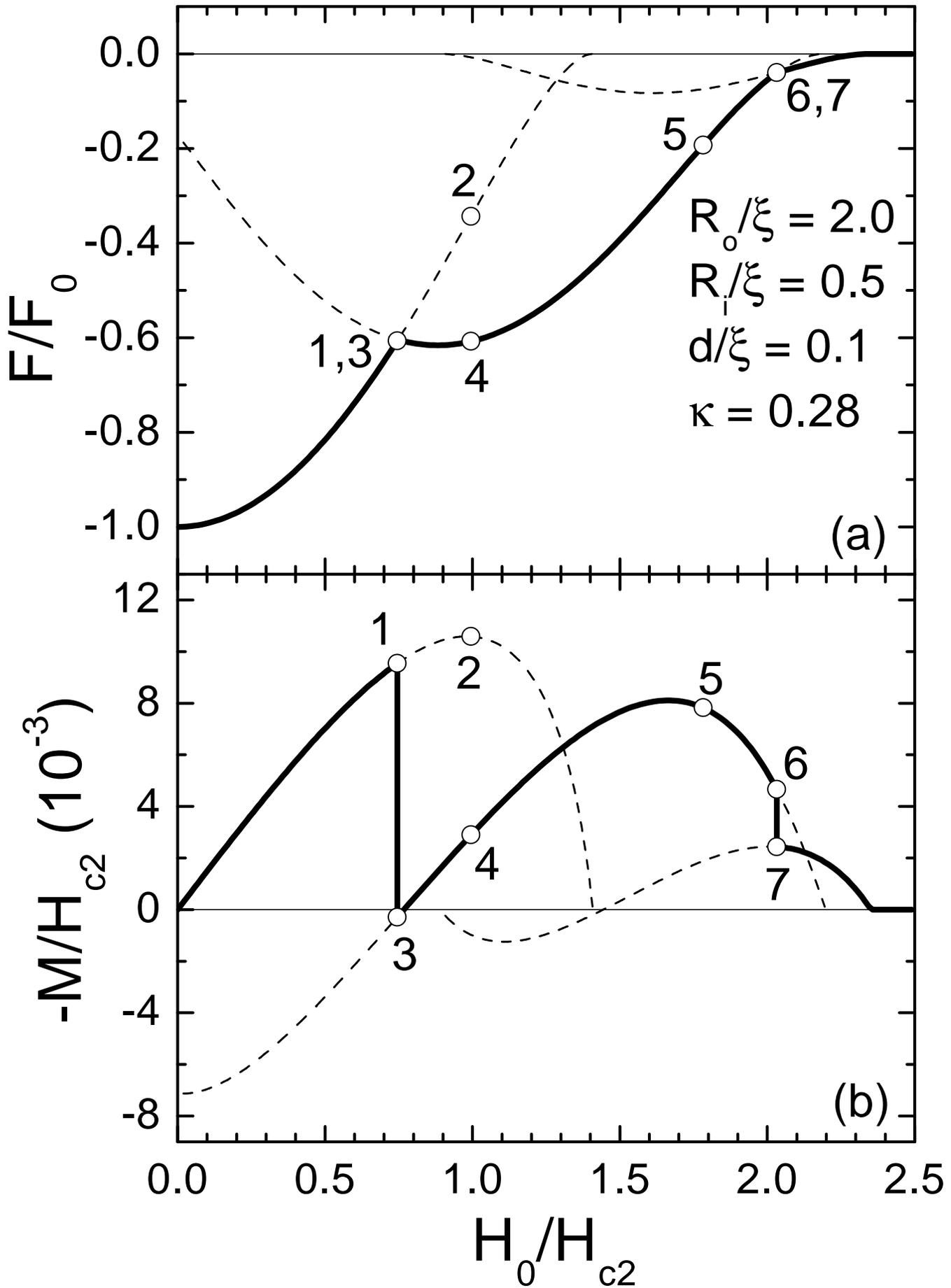
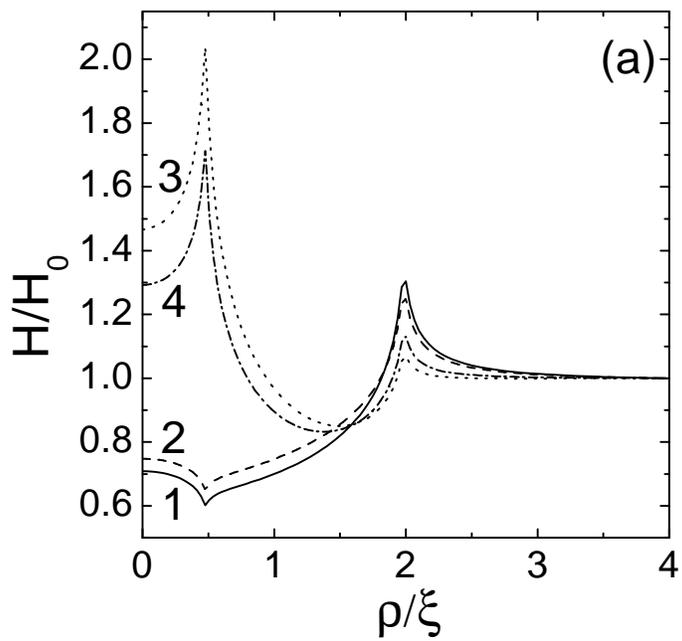
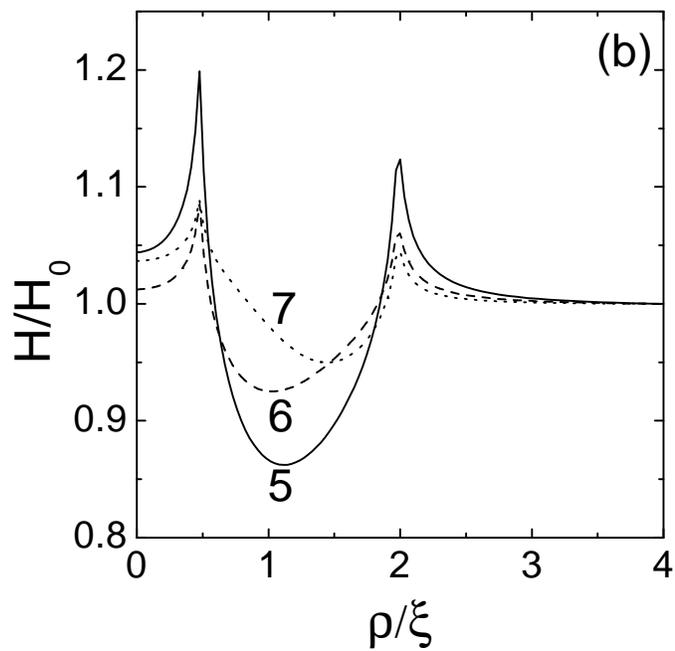
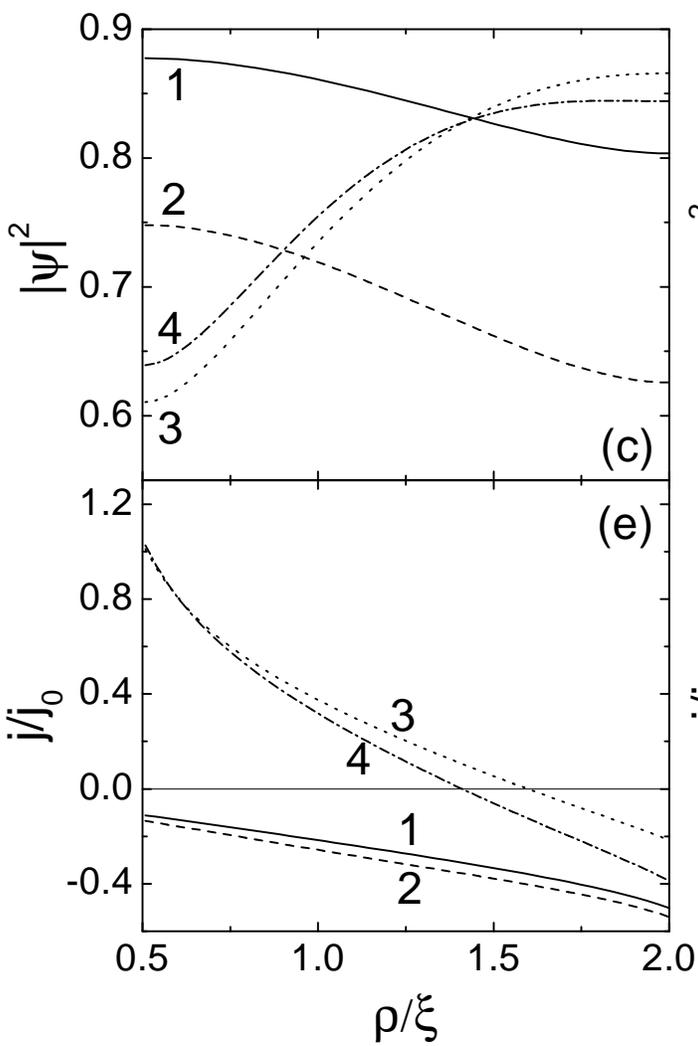
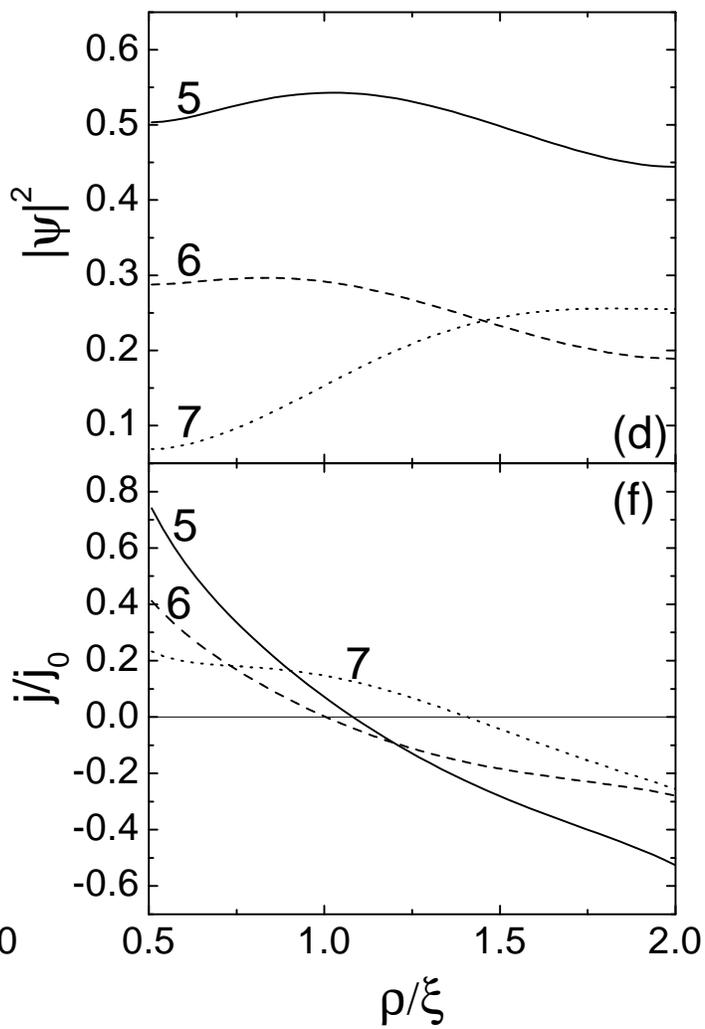
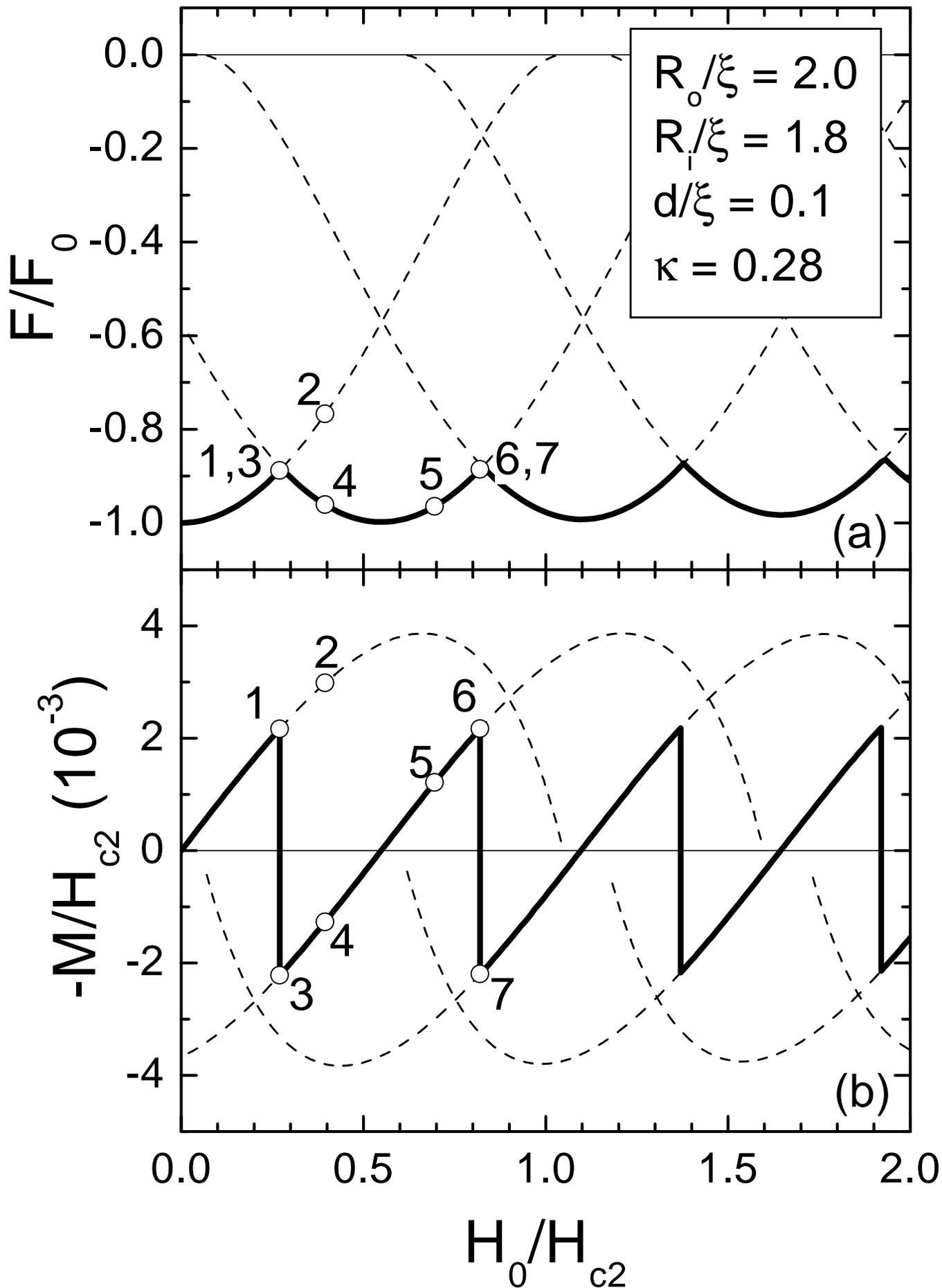
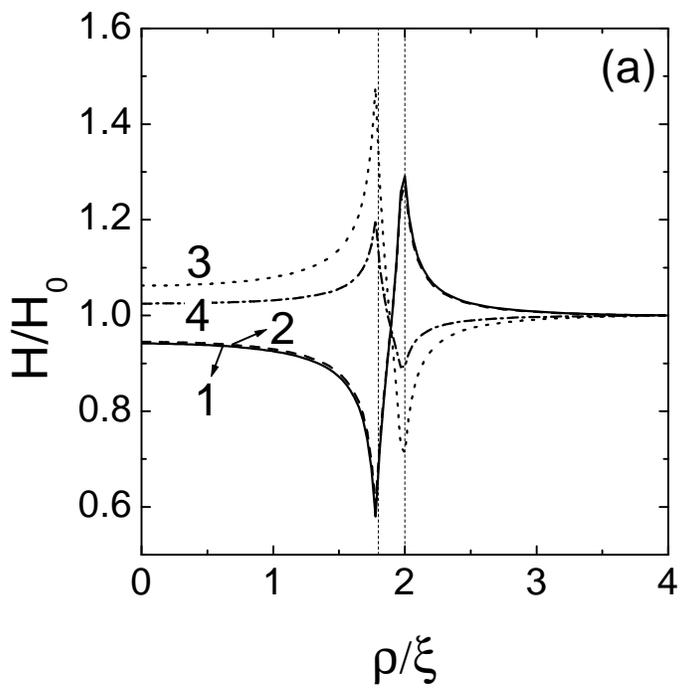
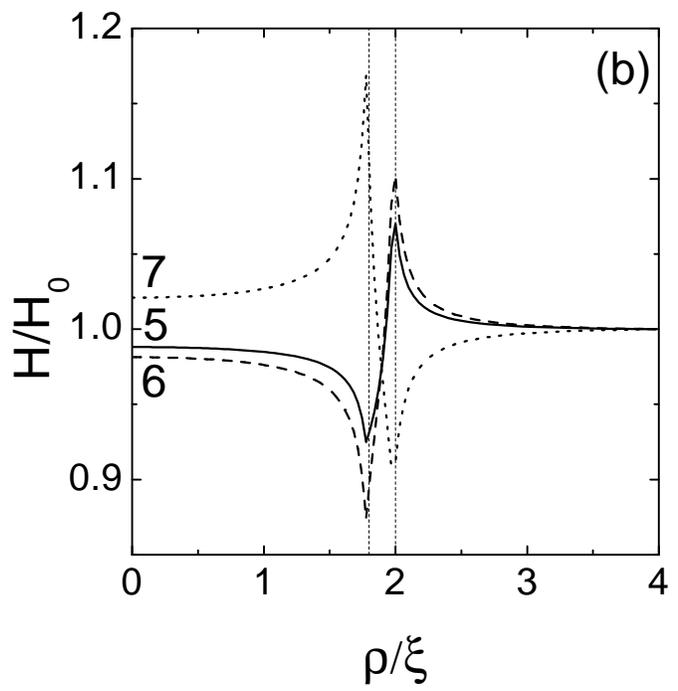
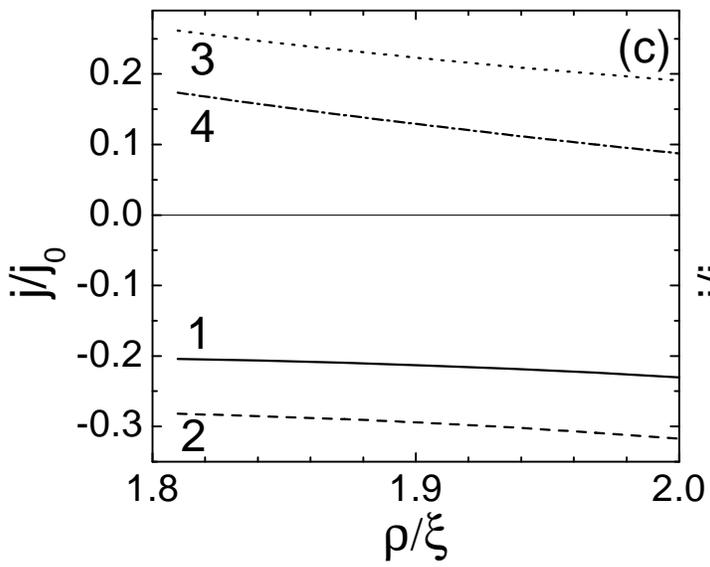
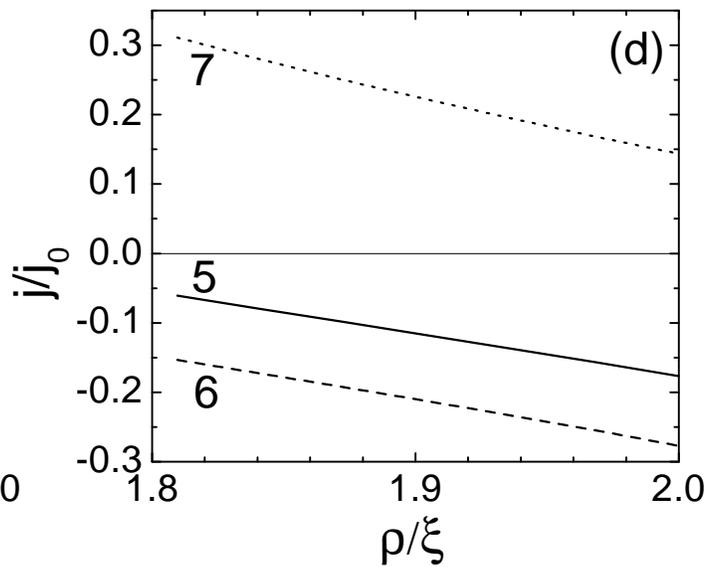
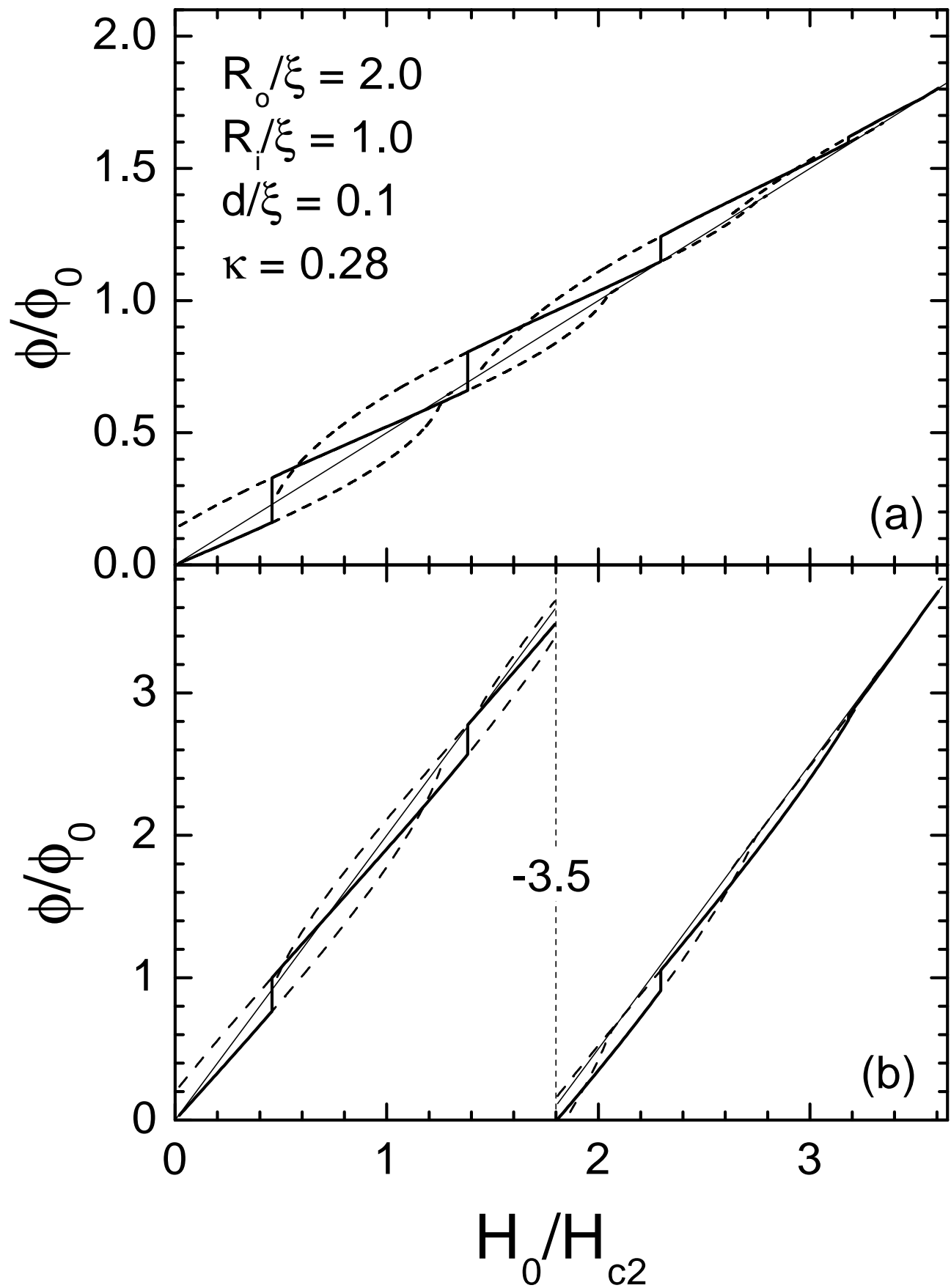
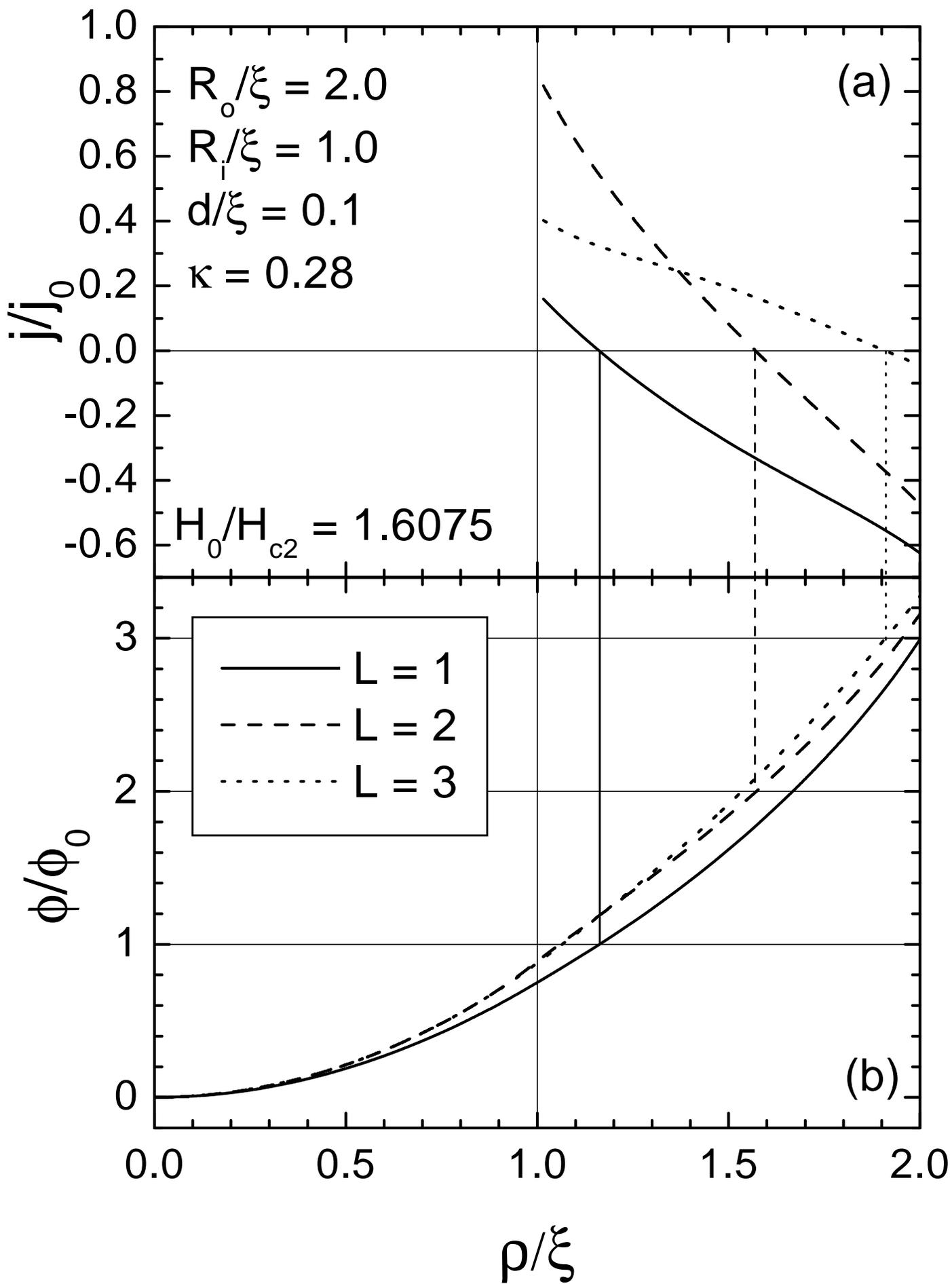
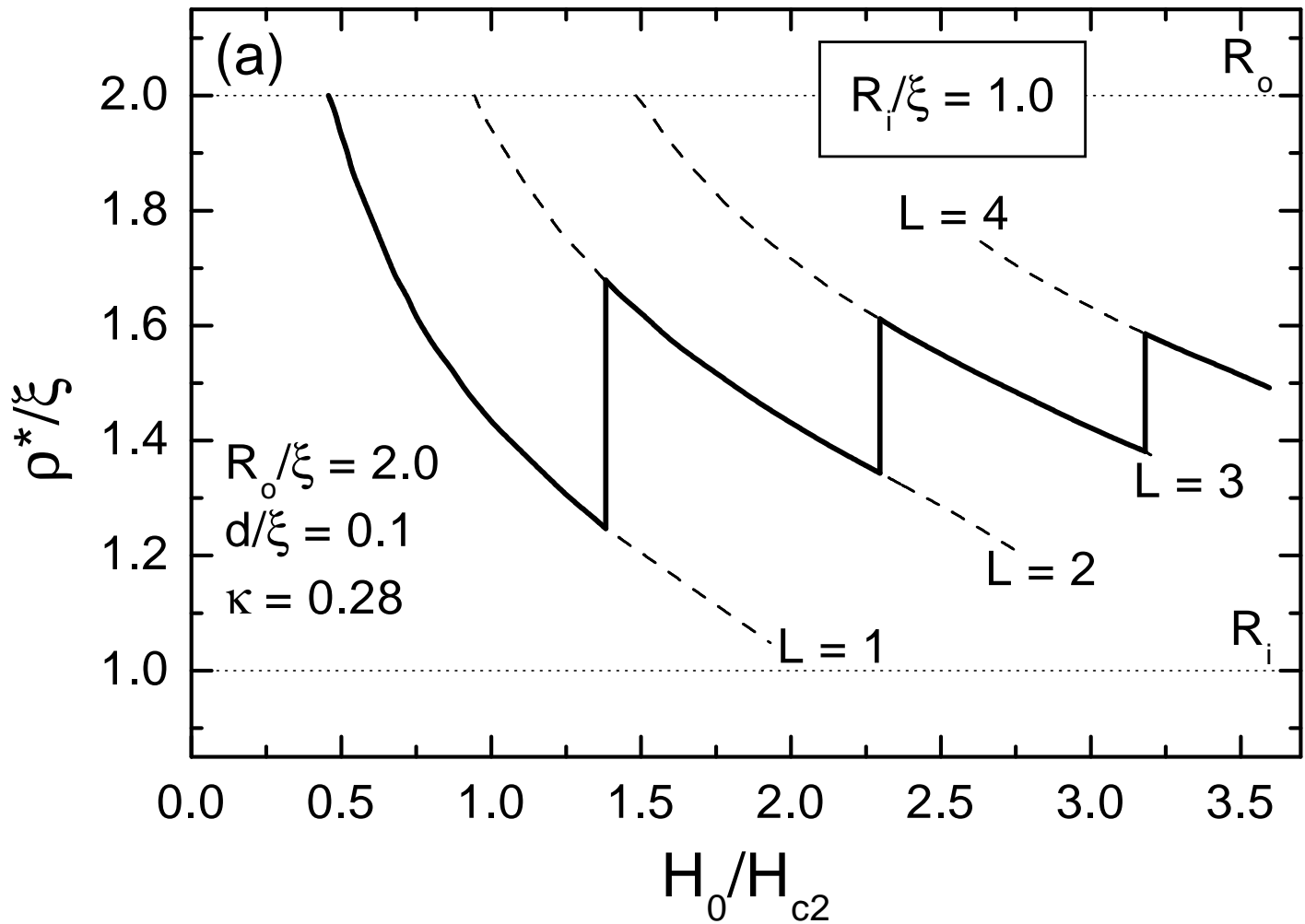
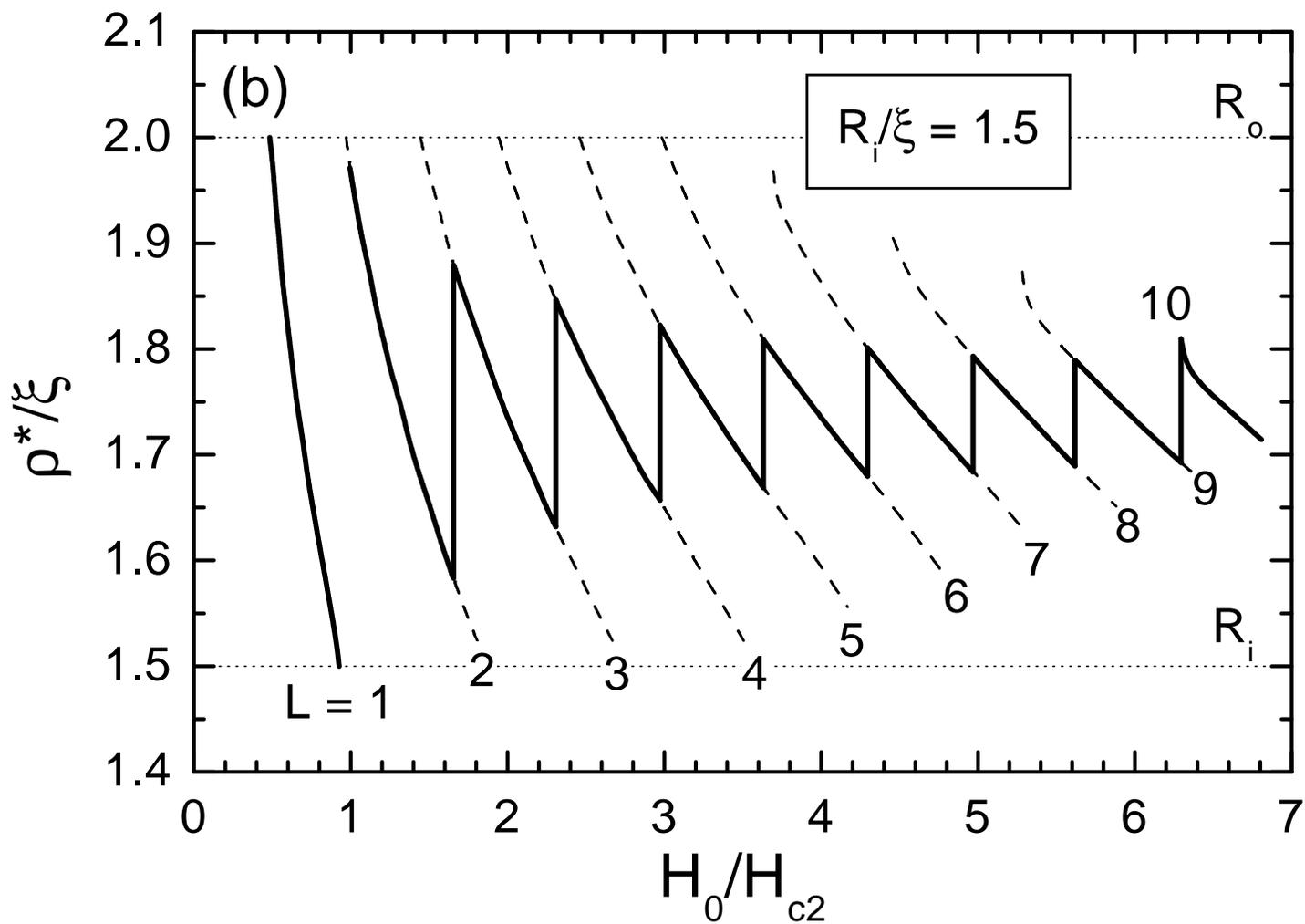
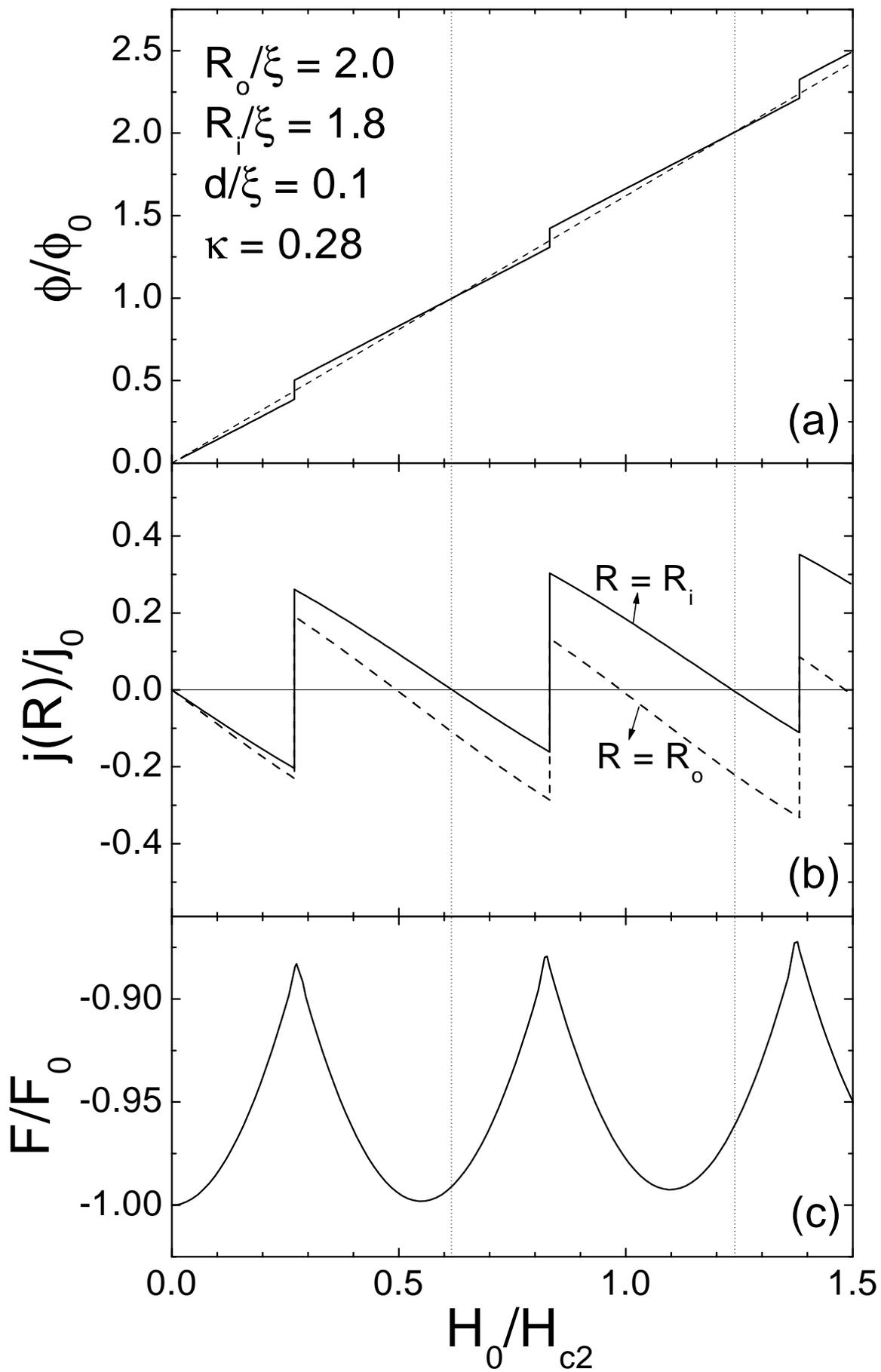
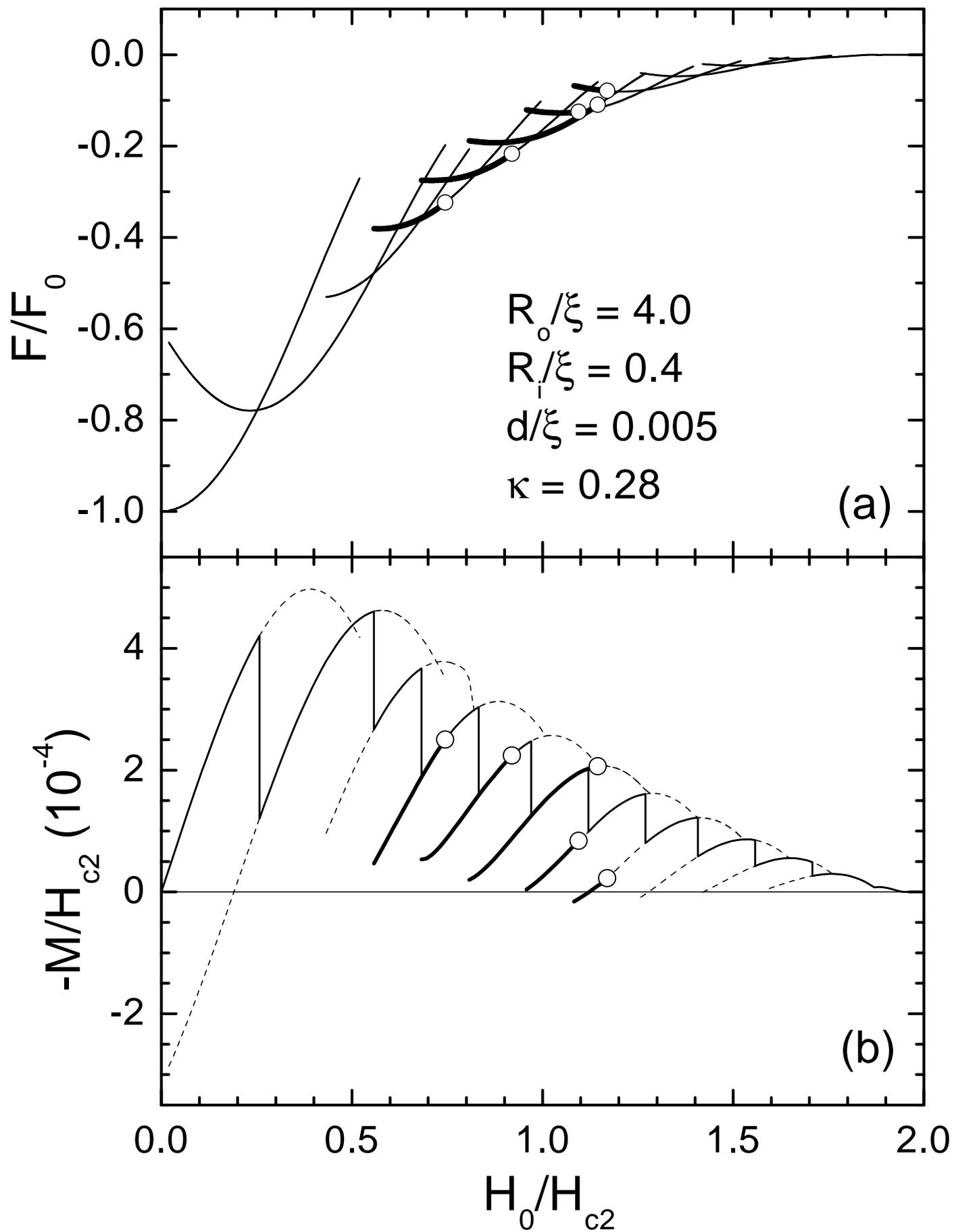
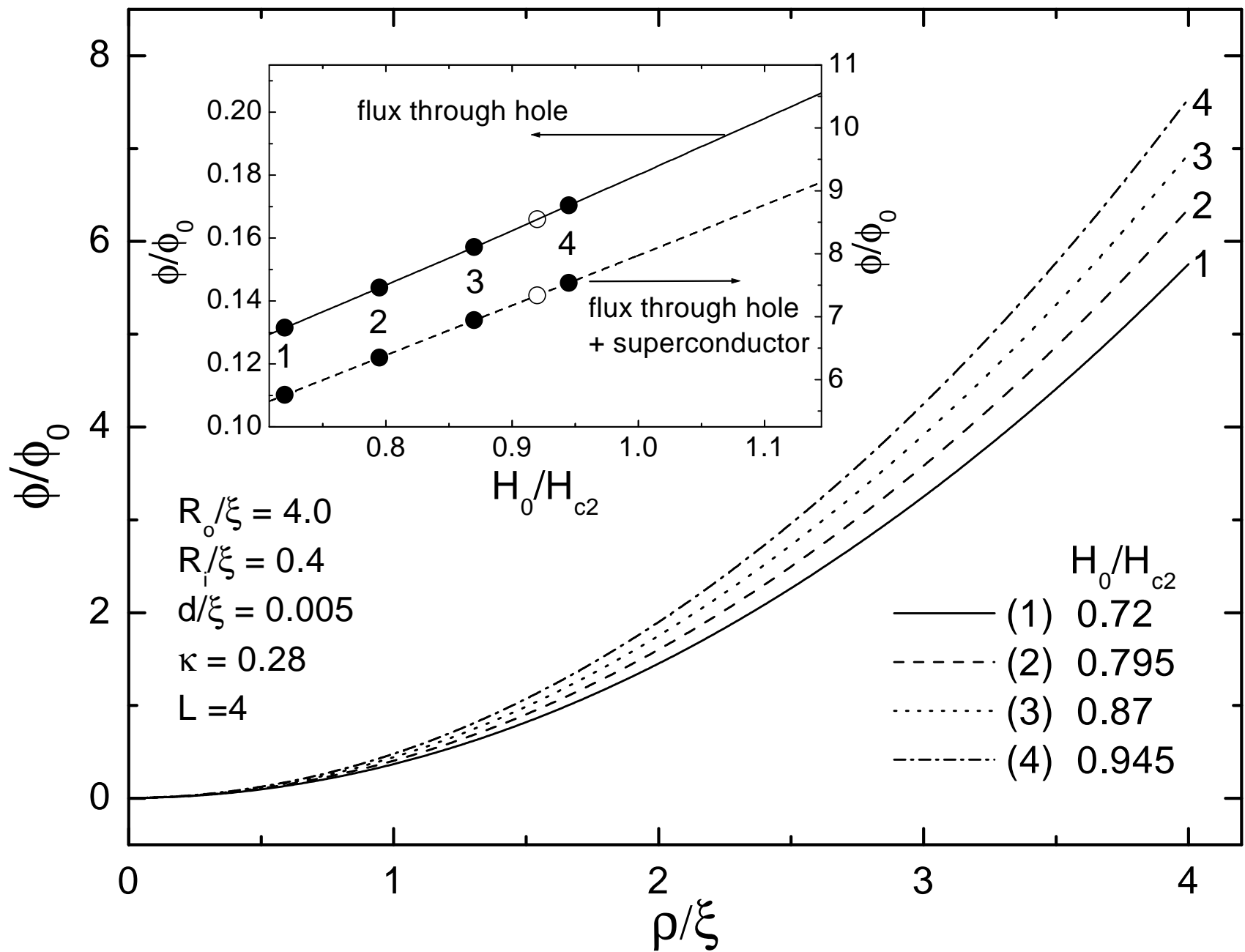
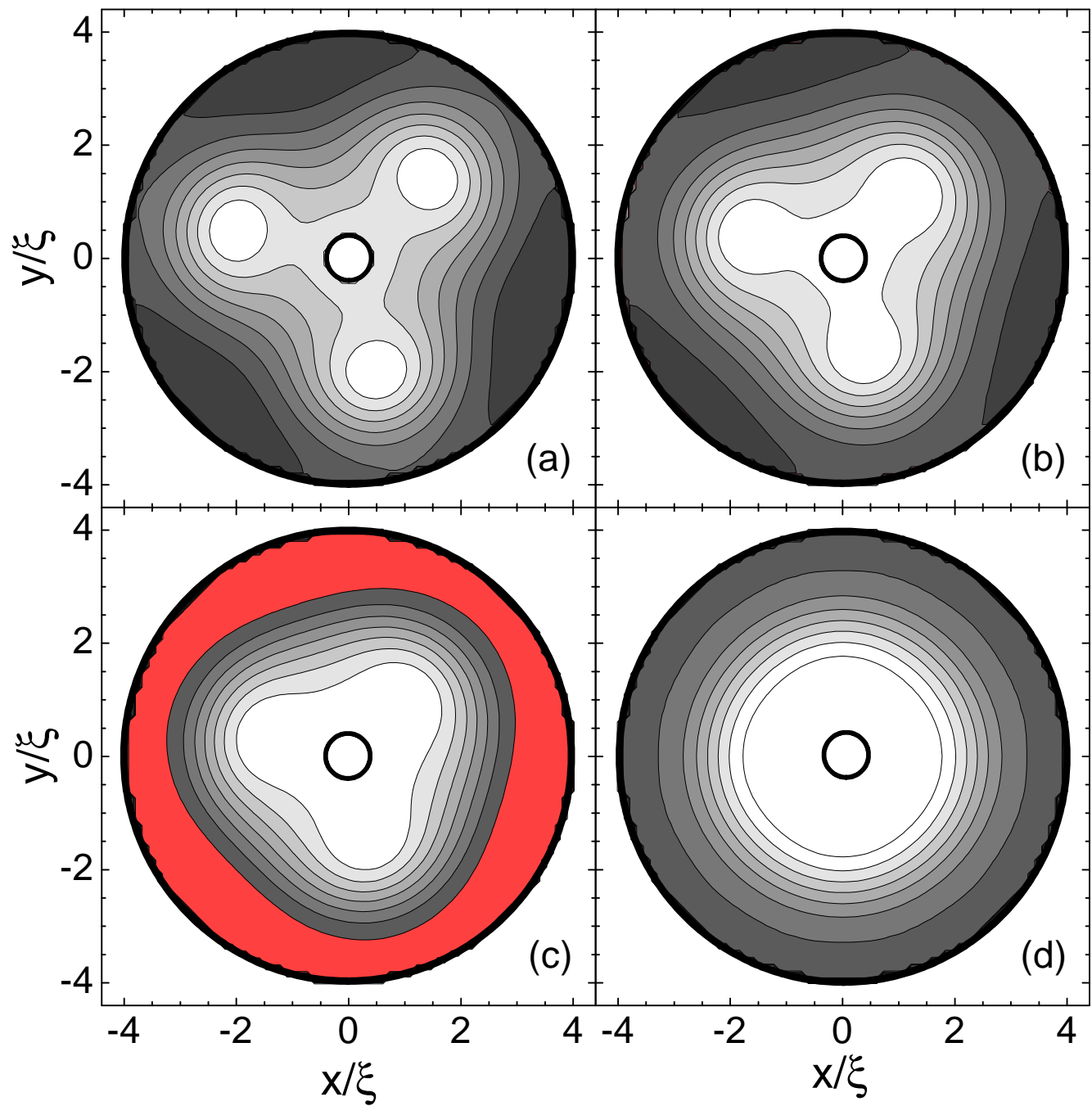
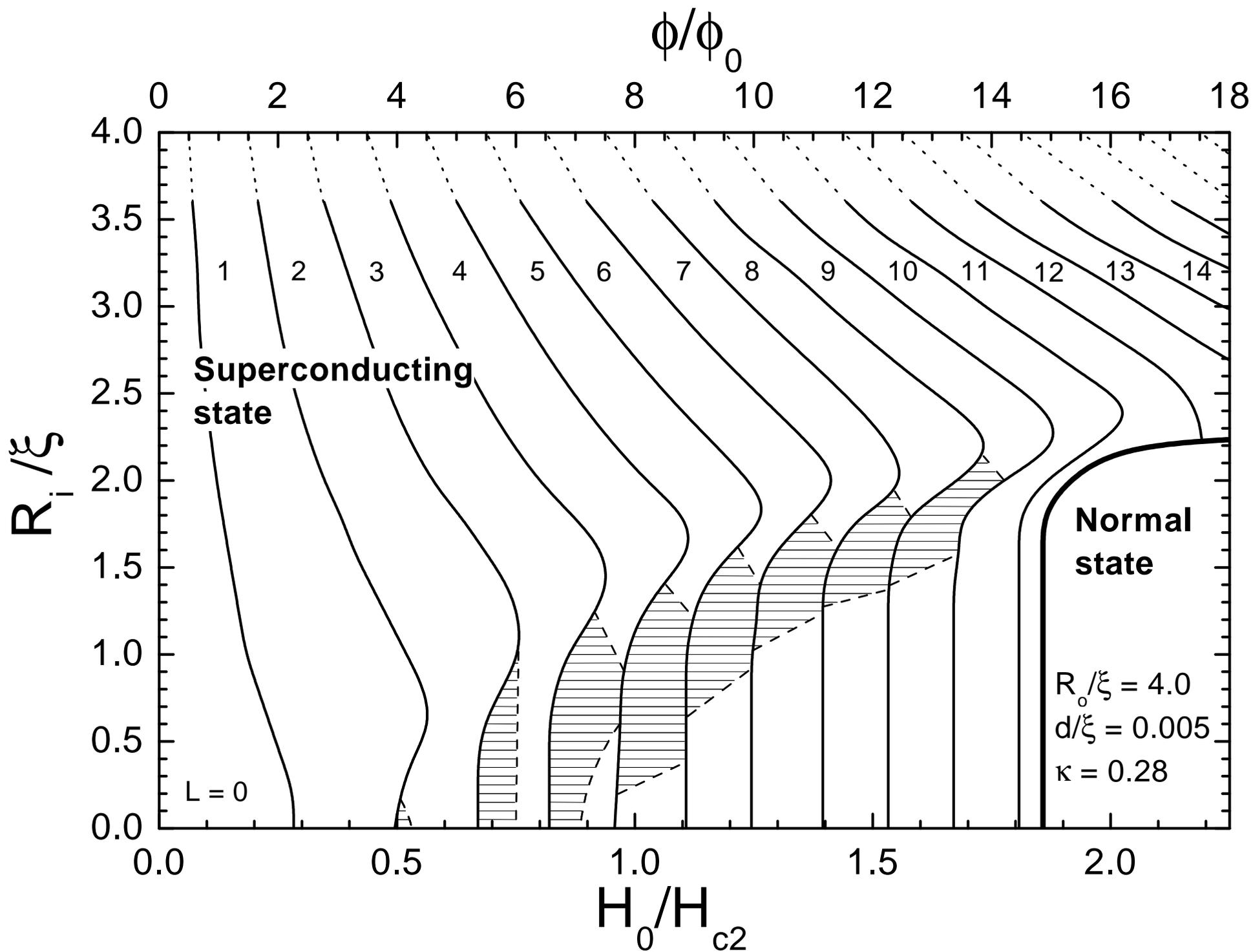
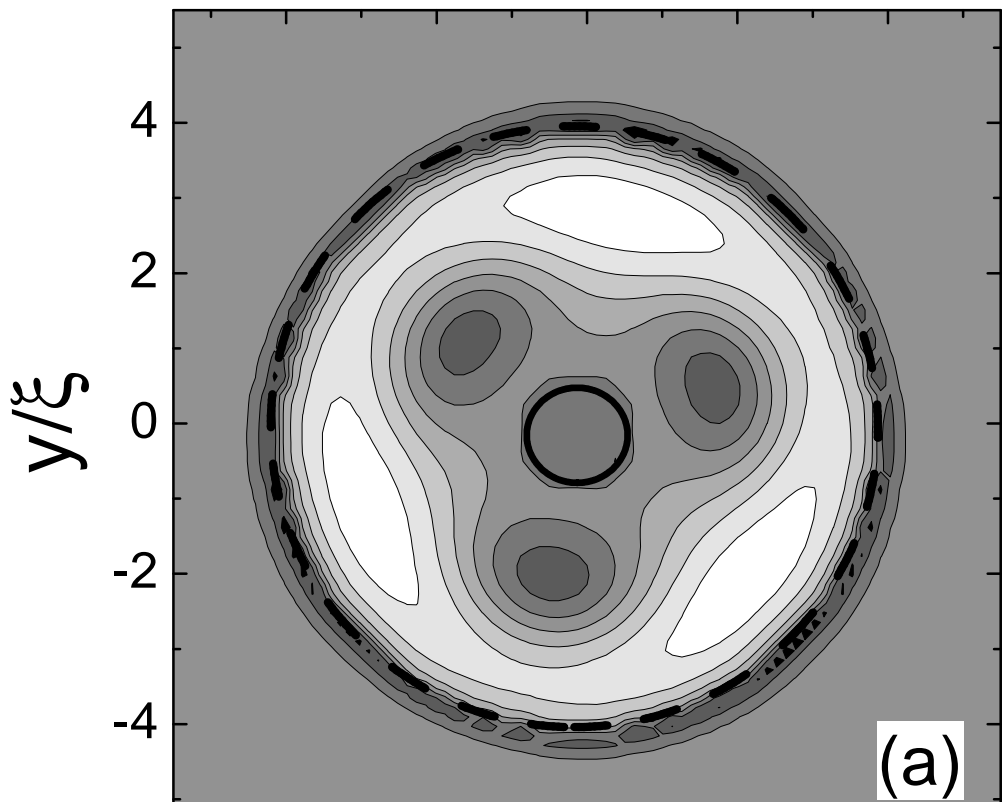
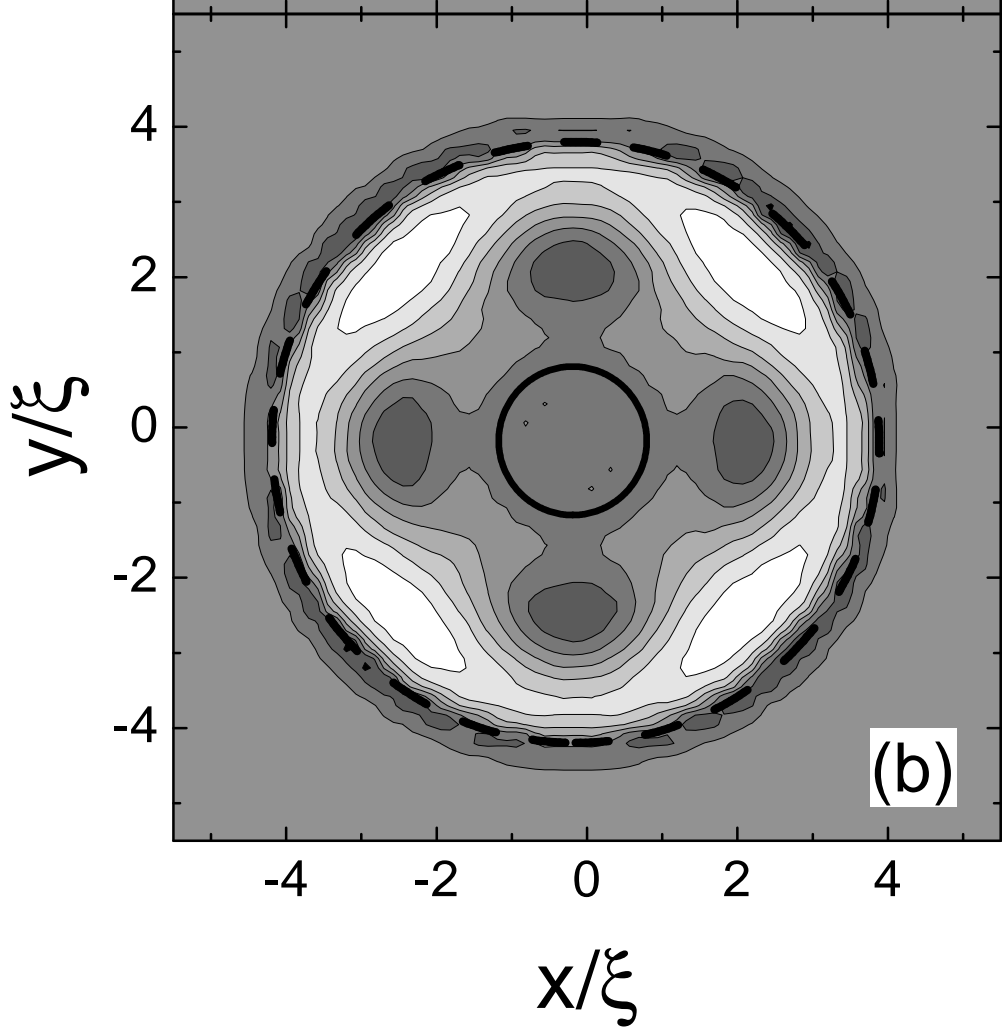
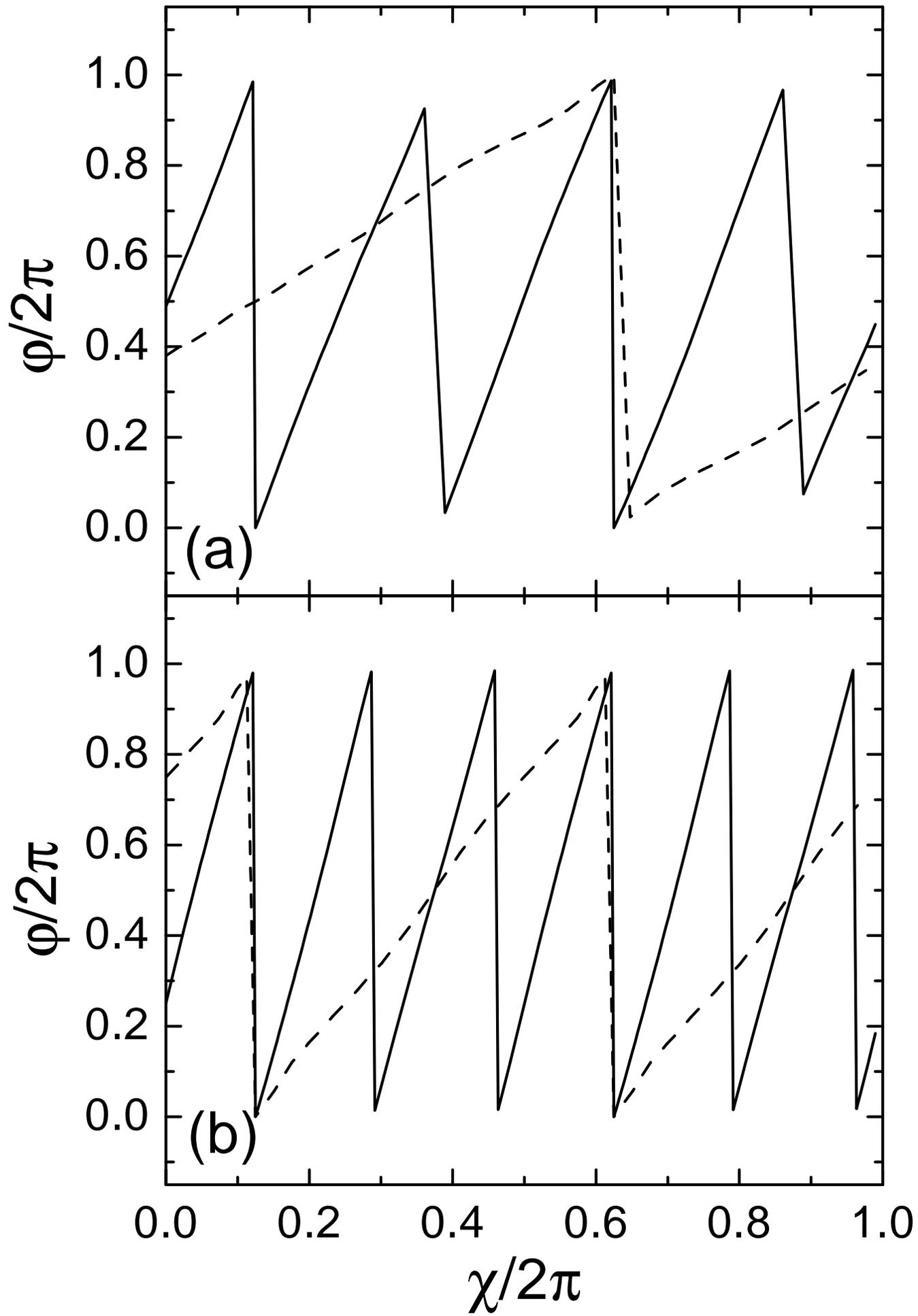
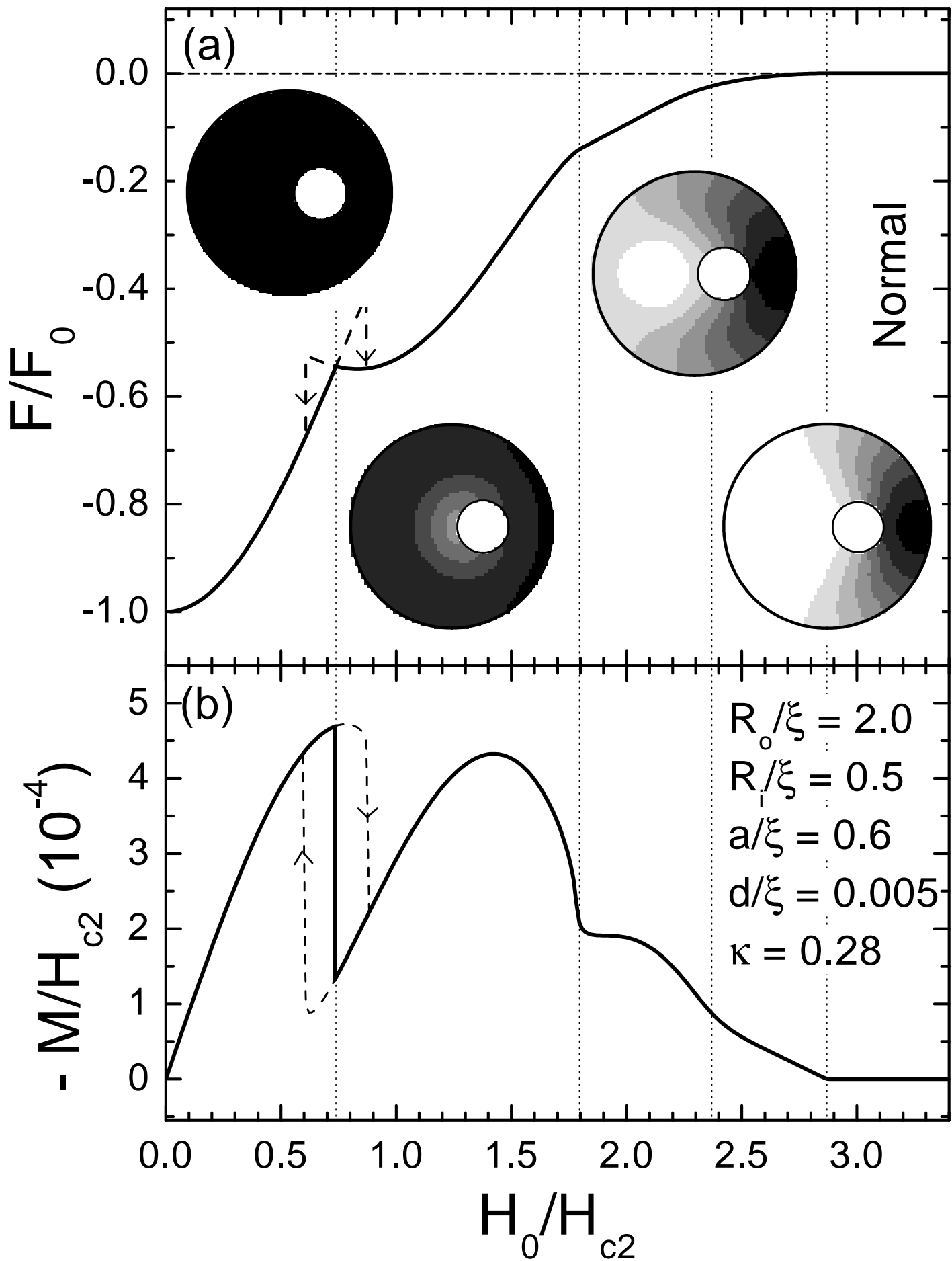
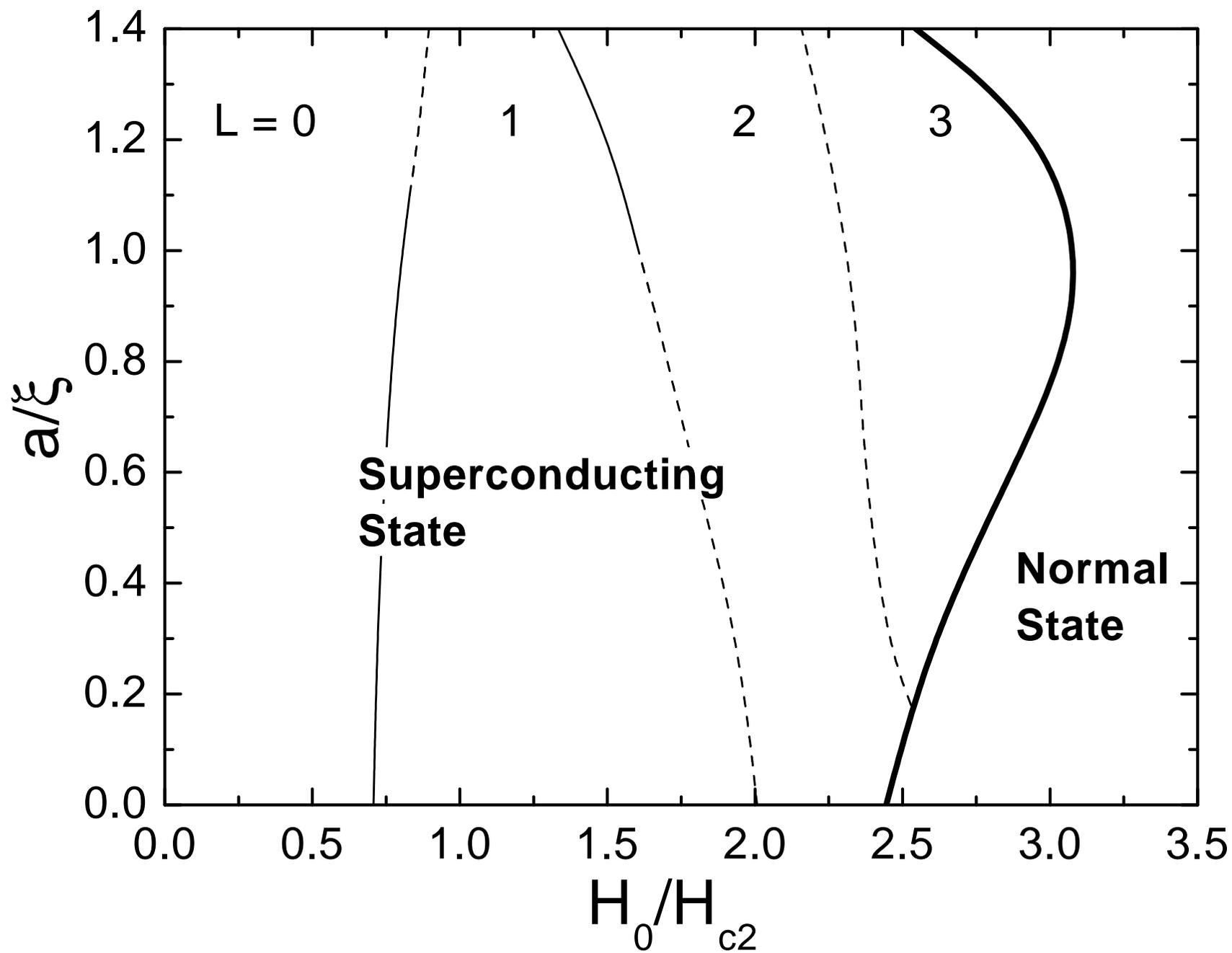
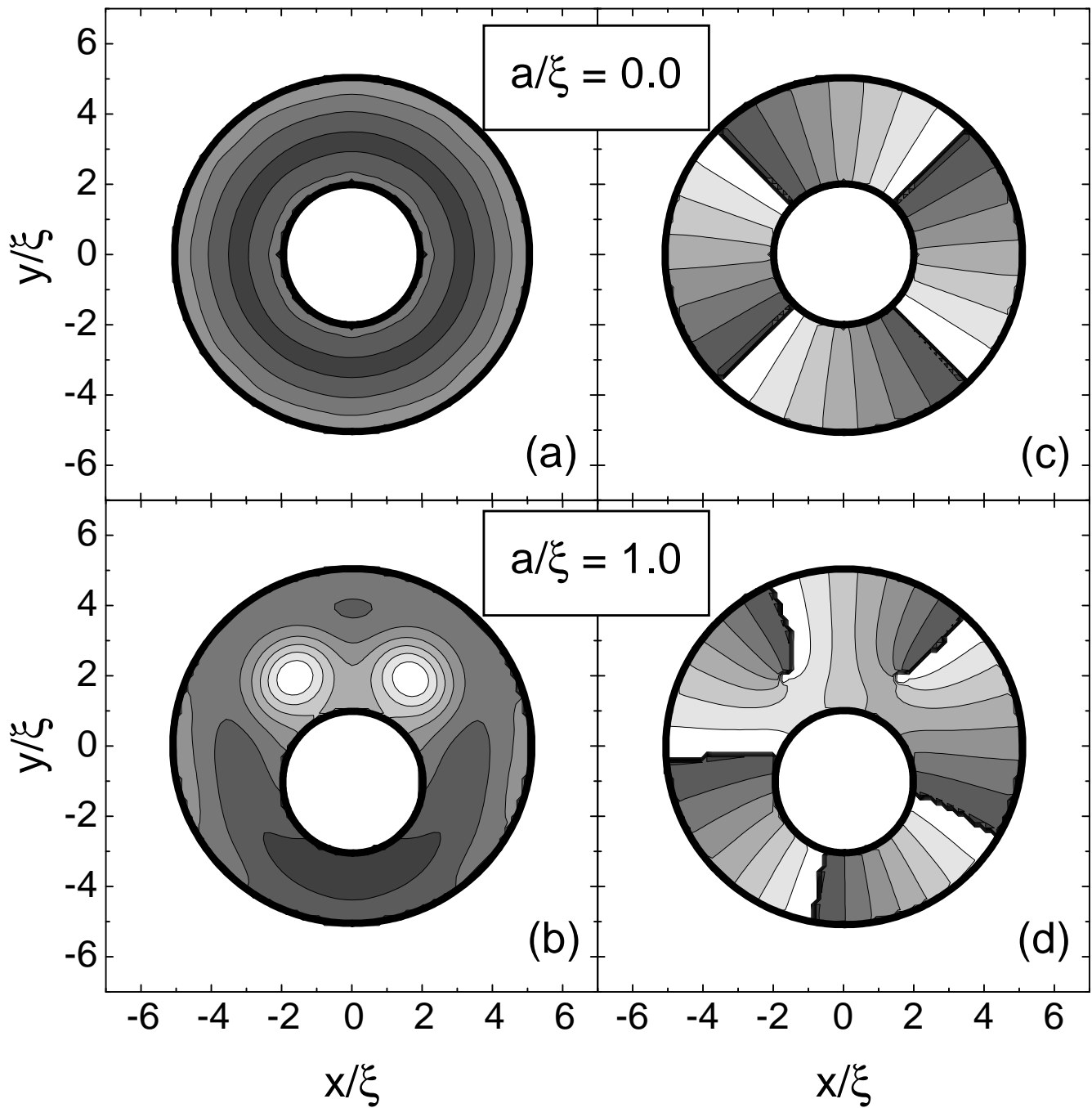
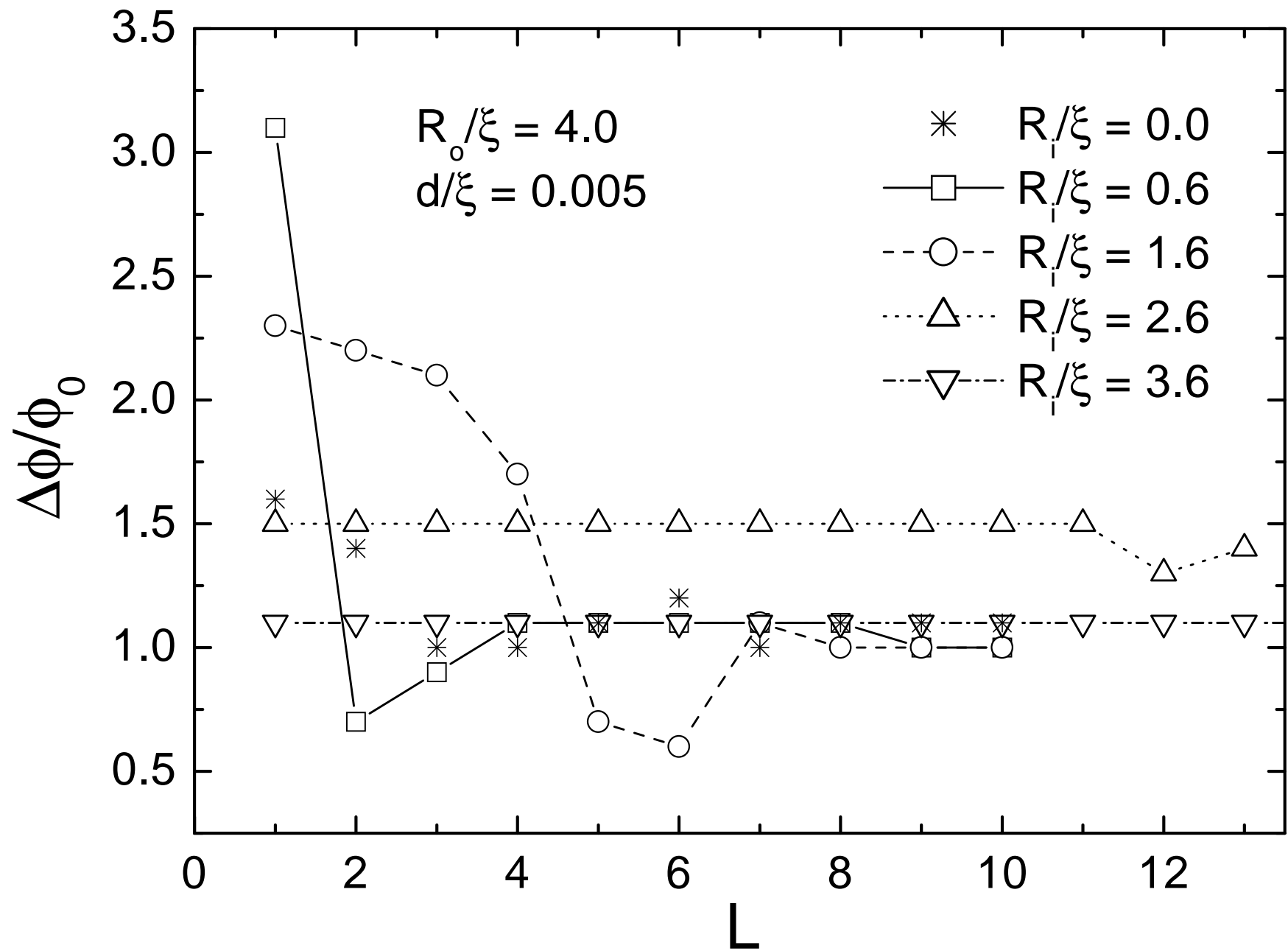
\end{document}